\documentclass[11pt]{article}
\usepackage{amsmath,amsthm,latexsym,amssymb,amsfonts,epsfig}

\newtheorem{lm}{Lemma}

\newtheorem{prop}[lm]{Proposition}
\newtheorem{prope}[lm]{Property}
\newtheorem{df}[lm]{Definition}
\newtheorem{thr}[lm]{Theorem}
\newtheorem{cor}[lm]{Corollary}

\numberwithin{equation}{section}
\numberwithin{lm}{section}
\newcommand{\R}{\mathbb{R}}                  % the reals
\newcommand{\C}{\mathbb{C}}                  % the complex numbers
\newcommand{\N}{\mathbb{N}}                  % the natural numbers

\newcommand{\ka}{{\bf k}}
\newcommand{\scpr}[2]{\left\langle\, #1\,,\, #2\, \right\rangle} % scalar prod.
\newcommand{\bra}[1]{|{\bf{#1}}\rangle}                        % bracket
\newcommand{\kwa}{\mathfrak{A}}           % algebra of fluxes and holonomie
\DeclareMathOperator{\cyl}{Cyl}
\newcommand{\Cyl}{\cyl}               % Cyl
\newcommand{\hilb}{\mathcal{H}}               % representation Hilbert-space
\newcommand{\one}{\text{\bf 1}}               % identity operator
\newcommand{\Si}{{\cal S}}
\newcommand{\F}{{\cal F}}
\newcommand{\FS}{{\cal F}_{\Si}}
\newcommand{\FSC}{{{\cal F}_{\Si}}^\C}

\newcommand{\gw}[1]{{#1}^\ast}

\begin{document}
\title{Background independent quantizations: the scalar field II}
%{Kinematical quantum algebra for theories of connections and
%uniqueness of its representation in the diffeomorphism invariant
%context}
\author{Wojciech Kami\'nski$^1$\thanks{wkaminsk@fuw.edu.pl},
        Jerzy Lewandowski$^{1,2}$\thanks{lewand@fuw.edu.pl}, Andrzej
Oko{\l}\'ow$^{1,3}$\thanks{oko@fuw.edu.pl}}
\date{\it 1. Instytut Fizyki Teoretycznej,
Uniwersytet Warszawski, ul. Ho\.{z}a 69, 00-681 Warszawa, Poland\\
2. Perimeter Institute for Theoretical Physics, 31 Caroline Street North,
Waterloo, Ontario N2L 2Y5, Canada\\
3. Department of Physics and Astronomy, Louisiana State University,
Baton Rouge, LA 70803-4001
} \maketitle
\begin{abstract}
We are concerned with the issue of quantization of a scalar field in
a diffeomorphism invariant  manner. We apply the method used  in
Loop Quantum Gravity. It relies on the specific  choice of scalar
field variables referred to as the polymer variables. The
quantization, in our formulation, amounts to introducing the
`quantum' {\it polymer
*-star algebra} and looking for  positive linear functionals, called
states. Assumed in our paper homeomorphism invariance allows to
derive the complete class of the states. They are determined by the
homeomorphism invariant states defined on the CW-complex
$*$-algebra. The corresponding GNS representations of the polymer
$*$-algebra  and their self-adjoint extensions are derived, the
equivalence classes are found and invariant subspaces characterized.
In the preceding letter (the part I) we outlined those results.
Here, we present the technical details.
\end{abstract}

%***************************************************
\section{Introduction}
%***************************************************

%***************************************************
\subsection{Motivation}
%***************************************************

Einstein's theory of gravity coupled (or not) to a matter field is a
prominent example of a so called ``background independent'' theory.
The phrase ``background independent'' means that the theory is
defined on a bare manifold endowed with no geometry or affine
structure. In this case, it is natural to look for a corresponding
quantum theory that is also manifestly  background independent (see
\cite{smolin} for a profound discussion of that issue). It is
conceivable, that a background independent quantum theory can even
be derived from a background {\it dependent} framework. Another
possibility is, that the classical limit of a given quantum theory
(the classical GR for example) has more symmetries (the
diffeomorphisms) then the underlying  quantum theory. Therefore, we
are not in the position to claim that every background dependent
approach to a quantization of a background independent classical
theory is wrong. Nonetheless, the first thing one should do is to
try without introducing  extra structures.  Loop Quantum Gravity is
the only known example. That name comes from the idea
\cite{rs1,rs2,aa1} of using field variables labeled by loops,
typically non-intersecting but knotted. Later, embedded {\it graphs}
were found a correct tool \cite{al2,jb1,jl-sn} which gave rise to
the present form of the theory. Therefore, there were attempts to
promote the theory under a new name like ``Quantum Geometry''
\cite{jl1}-\cite{dp2} or ``Quantum Spin Dynamics''
\cite{tt3}-\cite{tt5}. LQG  \cite{lqg} is a canonical theory, it
relies on the $3+1$ decomposition of space-time into the `space' $M$
and `time' $\R$. It is invariant with respect to the diffeomorphisms
(piecewise analytic) of $M$. It concerns the space-time geometry and
its dynamics, as well as coupled matter fields \cite{tt6}. The
matter fields however, have to be quantized in a new way, consistent
with the LQG and background independent quantization. And this is the point we
want to focus on in this work. We are concerned here with the
background independent, canonical quantization of the scalar field.
The first example of a background independent quantization of the
scalar field was proposed by Thiemann in his pioneering work
\cite{tt6}. It was improved and analyzed by various authors
\cite{als}. In the current paper we systematically derive a broader
class of examples. We show that our class is complete if certain
simple assumptions, like the topological invariance, are made. In
fact we consider only the GNS representations (and their
self-adjoint extensions) defined by invariant states introduced on
the appropriate  $*$-algebra. Among the constructed representations
we identify: those which give essentially self adjoint operators,
the equivalency classes, the irreducible representations. Each of
them can be applied  to the scalar field interacting with the
quantum geometry in the frame of LQG. A brief discussion of our
results is already published in the form of a letter \cite{klb}.

Here is an outline of the work.

In Sec. \ref{sec:pol-alg} we introduce the polymer $*$-algebra
$\kwa$, the key object studied in this work. It is constructed by
the canonical quantization of a classical field $\phi$ and the
canonically conjugate momentum $\pi$, both living on a manifold $M$.
The first step is choosing  basic field-variables. Briefly speaking
(see Sec. \ref{subsec:class-Lie} for the details) they are: the
`position variables' $e^{ik\phi(x)}$ (denoted $h_{k,x}$) and the
`momentum variables' $\int\pi f$ (denoted $\pi(f)$), where $k\in\R$
and $f$ is a smearing function. The vector space of the smearing
functions $\F$ is fixed in this section arbitrarily. Its specific
choice will play a crucial role later. The basic variables together
with the Poisson bracket form a Lie algebra, still classical. The
polymer $*$-algebra is defined (Sec. \ref{kwant}) by
`putting hats', meaning it is the quantum enveloping algebra of the
Lie algebra in question. It still depends on the choice of the
smearing function space $\F$. Next, we recall the definition of a
state on a $*$-algebra and elements of the GNS construction. The
conditions that will be imposed on the states are formulated in In
Secs. \ref{subsec:add} and \ref{subsec:sym}. Briefly, they are:
$(i)$ the diffeomorphism (Sec. \ref{sec:smooth}) and the
homeomorphism (Sec. \ref{sec:simplicial}-\ref{inv}) invariance, and
$(ii)$ an extra condition ensuring that for every point $x$, the
quantum operators $\rho(\hat{h}_{k,x})$, $k\in\R$, form a
1-dimensional group of unitary operators.

The first example of the polymer $*$-algebra considered in this work
is given by choosing for the smearing functions all the C$^{(n)}$,
$n=0,1,...,...\infty$, compactly supported functions on the manifold
$M$ (Sec. \ref{sec:smooth}). In this case we show there is exactly
one C$^{(n)}$-diffeomorphisms (homeomorphism in the $n=0$ case)
invariant state on the polymer $*$-algebra $\kwa$. The proof is a
simplified version of the proof used in \cite{lost} in the case of
the holonomy-flux $*$-algebra. This state coincides with the one
used thus far in LQG \cite{tt6,als}.

The proof of that first result  is very sensitive on the
differentiability class of the smearing functions being equal or
greater then the differentiability of the diffeomorphisms.
 The question that arises naturally, is: how much the uniqueness
proved in Sec. \ref{sec:smooth} depends on those differentiability
assumptions? Another natural choice for the smearing functions -also
used in LQG- is the characteristic functions of bounded regions in
$M$. The study of this case is the subject of the main part of this
work.

Specifically, the space $\FS$ of the smearing functions  we consider
throughout  Secs. \ref{sec:simplicial} - \ref{sec:inv} is spanned by
the characteristic functions of  ball-like regions in the manifold
$M$. The manifold is endowed with a piecewise-analytic structure and
the regions are assumed to be piecewise-analytic simplexes (see
Appendix). First (Sec. \ref{sec:simplicial}), we study the abelian
algebra $\exp(\odot\FS)$ freely generated by the characteristic
functions, and its complexification called here the CW-complex
$*$-algebra. That algebra, well defined on its own,  can be
identified with the subalgebra of the $*$-polymer algebra generated
by all  the momentum variables. We find all the homeomorphism
invariant states on the CW-complex $*$-algebra. They can be labeled
by the states $\mu:\C[\tau]\rightarrow\C$ where $\C[\tau]$ is the
$*$-algebra of the  polynomials of one real variable.

Every homeomorphism invariant state on the CW-complex $*$-algebra
admits a natural extension to the polymer $*$-algebra $\kwa$ (Sec.
\ref{sec:char}). The resulting  state defined on $\kwa$ is also
homeomorphism invariant. In this way a 1-1 correspondence between
the states $\mu:\C[\tau]\rightarrow\C$ and all the homeomorphism
invariant states on $\kwa$ which satisfy Property \ref{property}
(mentioned above) was established.

In the remaining part of the work (Secs. \ref{explicit2} -
\ref{sec:inv}) we are concerned with the properties  of the GNS
representations  corresponding to the homeomorphism invariant states
on $\kwa$ derived in Sec. \ref{sec:simplicial} and their and
self-adjoint extensions. An explicit form of the GNS representation
defined by a state found in Sec. \ref{sec:simplicial} in derived in
Sec. \ref{explicit2}. The issue of the self-adjointness of the
resulting operators is studied in Sec \ref{s-a}. In particular, we
characterize all those representations which admit a unique
self-adjoint extension according to Schmudgen \cite{schmud}. The
equivalence problem for the self-adjoint extensions of our
representations is solved in Sec. \ref{sec:equiv}. The reducibility
issue is solved in Sec. \ref{sec:inv}. The results are summarized in
Sec. \ref{sec:disc}.

 The problem we studied is also relevant for the issue of the uniqueness of the
diffeomorphism invariant state of quantum geometry in LQG. We
elaborate on that in Sec. \ref{sec:disc}.

%***************************************************

%***************************************************
\section{The polymer *-algebra $\kwa$ and GNS construction}
\label{sec:pol-alg}
%***************************************************

%***************************************************
\subsection{The classical scalar field}\label{subsec:class-phield}
%***************************************************

The classical scalar field in the canonical approach consists of a
pair of fields $(\phi, \pi)$ defined on a $N$-real dimensional
manifold $M$, where: $\phi\in C^{m_0}(M,\R)$ where $C^{m_0}(M,\R)$
stands for the space of the C$^{m_0}$ real valued functions
defined on $M$, $m_0\in\N$ or $m_0=\infty$, whereas $\pi$, called
the canonical momentum, is a scalar density of the weight $1$. The
momentum  $\pi$ can be expressed by  a function
$\tilde{\pi}:U\rightarrow \R$ defined on an arbitrary region
$U\subset M$ equipped with coordinates $(x^1,...,x^N)$. The
function, however, depends on the coordinates in the following
way:  if $({x'}^1,...,{x'}^N)$ is another coordinate system
defined on $U$, then the corresponding momentum function
$\tilde{\pi}'$ is such that at every $x\in U$
\begin{equation}
\tilde{\pi}(x)d^Nx\ =\ \tilde{\pi}'(x)d^Nx'.
\end{equation}

The fields $\phi$ and $\pi$ are called canonically conjugate in the sense
od the Poisson bracket,  usually wrote as
\begin{equation}\label{poisson}
\{\phi(x), \tilde{\pi}(y)\}\ =\ \delta_x(y).
\end{equation}

That Poisson bracket will be encoded  in the
Lie bracket of the Lie algebra introduced in the next
subsection.

%***************************************************
\subsection{The classical polymer Lie algebra}\label{subsec:class-Lie}
%***************************************************

\begin{df}\label{h}
A polymer  position variable is
a(n exponentiated evaluation)
function $h_{k,x}:C^{m_0}(M,\R)\rightarrow \C$
assigned to a pair  $(k,x)\in
\R\times M$ such that $k\not=0$, and defined as follows
\begin{equation}
h_{k,x}:\phi\mapsto  e^{ik\phi(x)}.
\end{equation}
\end{df}

The set of the polymer position variables is closed with respect
to the complex conjugation
 \begin{equation}
\overline{h_{k_i,x_i}}\ =\
h_{-k_i,x_i}.
\end{equation}
\begin{df}
The polymer position variable space $\cyl^\R_{1}$ is the set of all the real
valued linear combinations of all the polymer position variables,
that is the set of all the linear combinations
$h$ such that can be written in the form
\begin{equation}
h\ =\ \sum_{i=1}^n\left(a_ih_{k_i,x_i} + \overline{a_i}h_{-k_i,x_i}\right)
\end{equation}
where $n\in \N$,  $x_1,\ldots,x_n\in M$, $k_1,\ldots,k_n\in\R\setminus\{0\}$,
$a_1,\ldots, a_n\in \C$ are  arbitrary.
\end{df}
We could well be using only real valued functions ${\rm Re}\,h_{k,x}$
and ${\rm Im}\,h_{k,x}$ up to this point.
The complexification will come with the quantization.

A  momentum variable will be assigned to a  function
$f:M\rightarrow \R$. It can be thought of as the integral $\int_M\pi f$.
However, we will identify it with the operator defined by the Poisson
brackets  $\{h_{k,x},\int_M\pi f\}$:

\begin{df}\label{pi}
A  momentum variable is  a linear map $\pi(f) :
 \cyl^\R_{1}\rightarrow  \cyl^\R_{1}$ and
defined by the following action on the polymer position variables
\begin{equation}\label{pif}
\pi(f)h_{k,x}\ =\ i\,t\,f(x) h_{t,x}.
\end{equation}
\end{df}

This definition is our  version of the Poisson bracket
(\ref{poisson}). We  fix a vector subspace ${\cal F}$ of the space
of all the real valued functions ${\rm Map}(M,\R)$ defined on $M$
and refer to it the space of the smearing functions. We consider
only the  momenta defined by the smearing functions $f\in {\cal F}$.

\begin{df}\label{Pi}
The momentum variable space  $\Pi_{\cal F}$ defined by a given
space ${\cal F}\subset {\rm Map}(M,\R)$ of the smearing functions is
the real vector space spanned by the linear maps  $\pi(f)$
(\ref{pif}) such that $f\in{\cal F}$. The vector space ${\cal F}$
itself is referred to as the smearing functions space, and its
elements as smearing functions.
\end{df}
Actually, the map
\begin{equation}\label{F2Pi}
{\cal F}\ni f\mapsto \pi(f)\in \Pi_{\cal F},
\end{equation}
is an isomorphism of the vector spaces.

Finally, we define:
\begin{df}
The polymer Lie algebra $(\kwa_{\rm cl},\, \{\cdot,\,\cdot\})$
corresponding to  a momentum variable space $\Pi_{\cal F}$ is the
direct sum vector space
\begin{equation}
\kwa_{{\rm cl}}\ :=\ \cyl^\R_{1}\oplus\Pi_{\cal F},
\end{equation}
equipped with the following Lie bracket $\{\cdot,\,\cdot\}$:

\begin{equation}
\{(h,\pi),(h',\pi')\}\ :=\ (\pi'h-\pi h',0).
\end{equation}
\end{df}

The Lie bracket $\{\,\cdot,\,\cdot\}$ encodes the structure
of the Poisson bracket (\ref{poisson}) and this is the reason
we denote it in this way (rather then  the usual `$[\,\cdot,\,\cdot]$').

Later in this paper, we will consider the following two examples
of the smearing functions space ${\cal F}$ and the corresponding
of the momentum variable spaces:
\medskip

\noindent{\bf Example:}
\begin{enumerate}
\item
$${\cal F}=C^{m_0}_0(M,\R)$$
 where the second subscript $0$ stands for `compactly
supported'.
\item
${\cal F}$ spanned by the characteristic functions of suitably
defined family of open subsets of $M$
\end{enumerate}
Our main results will concern the second example. But until  the end
of this section we  keep the smearing functions space ${\cal F}$
general.

%***************************************************
\subsection{The quantum  polymer *-algebra $\kwa$ \label{kwant} }
%***************************************************
In order to turn the polymer  position  and momentum variables
(Definition \ref{h} and \ref{pi}) into quantum operators we will construct
from them a *-algebra. Briefly speaking, the commutation relations
of the operators will be defined as $i$ times the  Lie bracket
of the corresponding elements of the classical
Lie algebra. To do it in an exact way we formulate a definition
analogous to that of the universal enveloping algebra.

Let  $(\kwa_{\rm cl}, \{\cdot,\,\cdot\})$ be an arbitrary real Lie
algebra.

First, consider a complex Lie algebra
 $(\kwa_{\rm cl}^\C, \{\cdot,\,\cdot\})$, the complexification
of  $(\kwa_{\rm cl}, \{\cdot,\,\cdot\})$. There is the natural
complex conjugation $\bar{}:\kwa_{\rm cl}^\C\rightarrow
\kwa_{\rm cl}^\C$.

Next, consider the huge space
\begin{equation}
\bigoplus_{n=0}^{\infty} (\kwa_{\rm cl}^\C)^{\otimes n}
\end{equation}
where
\begin{equation}
(\kwa_{\rm cl}^\C)^0:=\C.
\end{equation}
It has  the natural complex associative algebra structure
defined by the complex vector space structure and the operation
$\otimes$.
There is also naturally defined anti-isomorphism-involution
$\gw{\cdot}$ in it, such that
\begin{equation}
\gw{(a_1\otimes\cdots\otimes a_n)}=\overline
{a_n}\otimes\cdots\otimes \overline{a_1}.
\end{equation}

Next, introduce the double sided ideal $J$ generated by the
following subset
\begin{equation}
\left\{\,a\otimes b- b\otimes a-i\{a,b\}\, :\, a,b\in \kwa_{\rm cl}^\C\,
\right\}.
\end{equation}
Ideal $J$ is preserved by the involution $\gw{\cdot}$, hence
$\gw{\cdot}$ passes to the quotient $\bigoplus_{n=0}^{\infty}
(\kwa_{\rm cl}^\C)^{\otimes n}/J$ (and is denoted by the same symbol
$\gw{\cdot}$).

Finally, define:

\begin{df}
The quantum enveloping algebra $(\kwa, \gw{\cdot})$ of a real Lie algebra
$(\kwa_{\rm cl}, \{\cdot,\,\cdot\})$ is the associative, unital,  *-algebra
\begin{equation}
\kwa\ =\ \bigoplus_{n=0}^{\infty} (\kwa_{\rm cl}^\C)^{\otimes n}/J.
\end{equation}
\label{kwaa}
\end{df}

From now on, until the end of the paper we will be considering the
quantum enveloping algebra of the polymer Lie algebra:
\begin{df}\label{poly*} A  polymer *-algebra $\kwa$ is
the quantum enveloping algebra of a polymer Lie algebra.
\end{df}
Elements of $\kwa$ can be heuristically thought of as quantum
operators corresponding to polynomials in the polymer position and momentum
variables. (They  become operators indeed, only given a representation
of $\kwa$ - see the next subsection).

We will be using the following notation:
an element $[a_1\otimes\dots\otimes a_n]\in \kwa$ corresponding
to $a_1\otimes\dots\otimes a_n$, where
$a_1,\dots,a_n\in \kwa_{{\rm cl}}^\C$ will be denoted by
 $\widehat{a_1}\dots \widehat{a_n}$,
\begin{equation}
\widehat{a_1}\dots \widehat{a_n}\ :=\ [a_1\otimes\dots\otimes a_n].
\end{equation}
In particular, we will be denoting
\begin{equation}
\hat{\pi}(f)\ =\ [\pi(f)]\in \kwa, \ \ \ \ \hat{h}_{k,x}=[h_{k,x}]\in\kwa,
\end{equation}
and will be calling them the quantum momentum, and, respectively,
quantum position variable.

The basic commutators $[A,B]=AB-BA$ in the polymer *-algebra $\kwa$ are
\begin{gather}
[\hat{h}_{k,x},\hat{\pi}(f)] = -kf(x)\hat{h}_{k,x},\label{com1}\\
[\hat{h}_{t,x},\hat{h}_{t',x'}]=0=[\hat{\pi}(f),\hat{\pi}(f)]\label{com2}
\end{gather}
for arbitrary $x,x'\in M$, $k,k'\in \R\setminus\{0\}$ and
$\pi(f)\in\Pi_{\cal F}$.

Owing to the commutation relations, every element of the polymer
*-algebra $\kwa$ corresponding to a momentum variable space $\Pi_{\cal F}$ can be
written as a complex linear combination of the elements of the form
\begin{equation}\label{prodhpi}
\prod_{i=1}^n\hat{h}_{k_i,x_i}\prod_{j=1}^m\hat{\pi}(f_j),
\end{equation}
where   $x_i \in M$,  $k_i\not=0$,
and $\pi(f_j) \in \Pi_{\cal F}$, and the
cases either $\prod_{i=1}^n\hat{h}_{k_i,x_i}$,  or
$\prod_{j=1}^m\hat{\pi}(f_j)$
are included as corresponding to $m=0$ and, respectively, $n=0$.

This decomposition is not necessarily unique, though. To begin with
the factors $h_{t_i,x_i}$ commute among themeselves as well
as the factors $\pi(f_i)$. The second source of the non-uniqueness,
is the identities satisfied by the momentum variables Definition \ref{pi}
\begin{equation}\label{idpi}
\pi(a_1f_1 + \ldots + a_kf_k)\ =\ a_1\pi(f_1) + \ldots a_k\pi(f_k)
\end{equation}
for every finite set of functions $f_1,\ldots,f_k\in {\cal F}$
and real numbers $a_1,\ldots,a_k$. On the other hand, there are
no identities in the vector space $\Phi$, that is
\begin{equation}
a_1h_{t_1,x_1}+\ldots+a_kh_{t_k,x_k}\ =\ 0\ \Rightarrow\ a_1=\ldots a_k =0,
\end{equation}
provided that $(k_i,x_i)\not= (k_j,x_j)$ for every $i\not= j$.

%***************************************************
\subsection{States on *-algebras and GNS representations}\label{subsec:GNS}
%***************************************************

This short subsection might  be skipped by the reader familiar with the
notion of states
on *-algebras and the GNS construction.

Quantization of the basic position and momentum variables for
the scalar field amounts to finding a representation of the
quantum algebra $\kwa$ of the basic variables in the space of operators
defined in a Hilbert space. In this paper we will be concerned with
the so called GNS (Gel'fand-Naimark-Segal) representations defined
by non-negative linear functional  on the *-algebra $\kwa$.

Recall, that:
\begin{df} \label{state} A state on a unital $*$-algebra $\kwa$
is a non-trivial linear functional
$\omega:\kwa\rightarrow\C$, such that for every $\hat{a}\in\kwa$
\begin{equation}\label{state1}
\omega(\gw{{\hat a}}{\hat a})\ \ge 0, \ \ \ \  {\rm and}\ \ \ \ \omega(\hat{1})=1,
\end{equation}
where $\hat{1}$ stands for the unity element.
\end{df}

In fact, if we do not assume the equality above, it follows from the
inequality (the positivity condition) itself, that
\begin{equation}
\omega(\gw{{\hat a}})\ =\ \overline{\omega({\hat a})}, \ \ \ {\rm and} \ \ \
\omega(\hat{1})\ >\ 0.
\end{equation}
Therefore, given the positivity, by rescalling  we can always achieve
the equality in Definition \ref{state}.

Given a state on $\kwa$, we  construct the corresponding
GNS representation
\[ \label{GNSrep}
(\hilb_\omega^0, \scpr{\cdot}{\cdot}_\omega, \rho_\omega,\Omega_\omega),
\]
where: $\hilb_\omega^0$ is a vector space endowed with a unitary
scalar product $\scpr{\cdot}{\cdot}_\omega$,   $\rho_\omega$ a
representation of $\kwa$ on $\hilb_\omega^0$, and $\Omega_\omega$ a
vector in $\hilb_\omega^0$ which, when viewed as a state on $\kwa$,
coincides with $\omega$ (see Equation \eqref{vec-state}). A detailed
exposition of the GNS construction for algebras of unbounded
operators can be found for example in \cite{schmud}. Here we will
only need the following elements and properties that are easy to
prove:

$(i)$ The vector space $\hilb_\omega^0$ is obtained as  the quotient
of $\kwa$ considered as a vector space
\begin{equation}\label{hilbomega}
%\left[\kwa \right]\
\hilb_\omega^0\ :=\ \kwa/{J_\omega}
\end{equation}
where the vector subspace
\begin{equation}
J_\omega:=\ \{{\hat a}\in\kwa\,:\, \omega(\gw{{\hat a}}{\hat a})=0\}
\end{equation}
 is a left ideal in $\kwa$;

$(ii)$ The unitary scalar product $\scpr{\cdot}{\cdot}_\omega$ is
defined in $\hilb_\omega^0$ by the state $\omega$, namely
\begin{equation}
 \scpr{[{\hat a}]}{[{\hat b}]}_\omega\ :=\ \omega(\gw{{\hat a}}{\hat b}),
\end{equation}
where for every ${\hat c}\in \kwa$, $[{\hat c}]\in \kwa/{J_\omega}$
stands for the equivalence class defined by $C$;
%$(ii)$  the
%product provides a norm $\|a\|_\omega=\sqrt{ \scpr{[{\hat a}]}{[{\hat a}]}}_\omega$ in
%$\left[ \kwa\right]$, and the completion
%\begin{equation}
%\hilb_\omega^0 :=\ \overline{\left[ \kwa \right]}
%\end{equation}
%together with the product $ \scpr{\cdot}{\cdot}_\omega$ is a Hilbert
%space;

$(iii)$ To every element $\hat a$ of $\kwa$ we assign a
linear (but in general unbounded) operator $\rho_\omega({\hat a})$,
defined in the entire vector space $\hilb_\omega^0$,
\begin{align}
\rho_\omega({\hat a}):\hilb_\omega^0&\rightarrow \hilb_\omega^0\\
\rho_\omega({\hat a})[{\hat b}]\ &:=\ [{\hat a}{\hat b}];\label{rhoGNS}
\end{align}

$(iv)$ The vector $\Omega_\omega$ is
\begin{equation}
\Omega_\omega\ =\ [\hat{1}],
\end{equation}
and its relation with the state $\omega$ is
\begin{equation}
\omega(\hat{a})\ =\
\scpr{\Omega_\omega}{\rho_\omega(\hat{a})\Omega_\omega}_\omega,
\label{vec-state}
\end{equation}
for every ${\hat a}\in \kwa$;

%$(i)$ the action $\rho_\omega$  preserves the subspace
%$\left[\kwa\right]$, hence $\left[\kwa\right]$
%serves as a common, dense domain for all the  operators
%$\rho_\omega(a),\ \ a\in\kwa$.

$(v)$ The  map $\rho_\omega:\kwa\rightarrow {\rm End}(\hilb_\omega^0)$
is an homomorphism of the associative algebra $\kwa$ into the
associative algebra  of the endomorphisms of the vector space
$\hilb_\omega^0$.

$(vi)$ The map $\rho_\omega$ is consistent with the *-structure of
$\kwa$ in the following way:
\begin{equation}\label{rhoA*}
 %\scpr{\rho_\omega({\hat a})^\dagger[{\hat b}]}{[{\hat c}]}_\omega\ =\
\scpr{\rho_\omega({\hat a}^*)[{\hat b}]}{[{\hat c}]}_\omega
\ =  \scpr{[{\hat b}]}{\rho_\omega({\hat a})[{\hat c}]}_\omega,
\end{equation}
for every ${\hat a},{\hat b},{\hat c}\in \kwa$.

A representation may be equivalent to a GNS representation.
For the sake of precision we spell out the definition of
equivalence:

\begin{df}\label{equiv}
Given  $(\,V,\,(\cdot|\cdot),\,\rho,\,v_0)$, where: $(a)$ $V$ is a vector  space  endowed with a
unitary form  $(\cdot|\cdot)$,
$(b)$ $\rho$ is an associative epimorphism
$$\rho: \kwa\rightarrow {\rm End}(V)$$
of a *-algebra $\kwa$  into the endomorphism algebra of $V$,
and $(c)$ $v_0\in V$, we say, that   $(\,V,\,(\cdot|\cdot),\,\rho,\,v_0)$
  is equivalent to a given GNS representation
$(\hilb_\omega^0, \scpr{\cdot}{\cdot}_\omega, \rho_\omega,\Omega_\omega)$
whenever there exists a unitary vector space isomorphism
${\cal I}:\hilb^0_\omega\rightarrow V$, such that
$$ \rho_\omega\ =\ {\cal I}^{-1} \circ \rho \circ {\cal I},\ \ \ \ {\rm and}
\ \ \ \ {\cal I} (\Omega_\omega)\ =\ v_0. $$
\end{df}
Notice, that the notion of the Hilbert space is not necessary
when one deals with the GNS algebras, as long as we are interested
in the elements of the algebra $\kwa$ only.

However, as for every unitary space, one can consider the Hilbert
space
\[
\hilb_\omega:=\overline{\hilb_\omega^0}.
\]
From this point of view, a difference between the current *-algebra
case and any other case of a $C^*$-algebra, is that since a general
operator $\rho_\omega({\hat a})$ is  unbounded, an equality
\begin{equation}\label{a*=a}
\gw{{\hat a}}={\hat a}
\end{equation}
 if it takes place, does not imply that
$\rho_\omega({\hat a})$ acting in the domain $\hilb^0_{\omega}$ is
essentially self-adjoint. If the algebra $\kwa$ is commutative then,
remarkably, (\ref{a*=a}) implies that the operator
$\rho_\omega(\hat{a})$ admits a self-adjoint extension. Indeed, the
operator commutes with the anti-linear operation
$$*:\hilb_\mu\rightarrow\hilb_\mu$$
induced  by the *-structure of $\kwa$, in the commutative case. In
general, the self-dual extension is not unique. However, due to the
algebra of the problem, this unambiguity does not lead to any
inconsistencies. Of course, given a *-algebra, it is always
interesting to ask which states define the GNS representations such
that the real elements of the algebra correspond to the
essentially-self adjoint operators.

%***************************************************
\subsection{Additional assumption about the representations of $\kwa$}
\label{subsec:add}
%***************************************************

In this paper, we apply the GNS construction to the polymer
*-algebra $\kwa$ (Definition \ref{poly*}).

Throughout  this paper, we will be assuming about a state $\omega$
on $\kwa$ that the corresponding GNS representation $\rho_\omega$
satisfies the following property:

\begin{prope}\label{property}
For every point $x\in M$, the set of operators
\begin{equation}
U_x\ =\ \{\rho_\omega(\hat{h}_{k,x})\,:\,k\in \R \}
\end{equation}
where $\hat{h}_{0,x}$ stands for the algebra unity element
$\hat{1}$, is a group (with the operation of the composition),
and
\begin{equation}\label{rhohh}
\rho_\omega(\hat{h}_{k,x})\rho_\omega(\hat{h}_{k',x})=
 \rho_\omega(\hat{h}_{k+k',x}).
\end{equation}
\end{prope}

This assumption reflects the property of the
exponentiated evaluation functions
\begin{equation}
e^{ik\phi(x)}e^{ik'\phi(x)}\ =\ e^{i(k+k')\phi(x)}.
\end{equation}
It is analogous to the property
\begin{equation}
\rho_\omega(\hat{\pi}(f_1))+\rho_\omega(\hat{\pi}(f_2))\ =\
\rho_\omega(\hat{\pi}(f_1+f_2)),
\end{equation}
already satisfied as an identity due to the definition of
the Lie algebra of the basic variables.

It follows from (\ref{rhohh}) that for every polymer position
variable $h_{k,x}$, the operator
\begin{equation}
\rho_\omega(\hat{h}_{k,x}): \hilb^0_\omega\rightarrow
\hilb^0_\omega
\end{equation}
 is the unitary form preserving and bijective, therefore it is
extendable by the continuity to a unitary operator in the Hilbert
space ${\hilb_\omega}$. Indeed, it follows that:
\begin{align}
\rho_\omega(\hat{h}_{-k,x})\rho_\omega(\hat{h}_{k,x})\ &=\
\rho(\hat{1})={\rm id}\ =\
\rho_\omega(\hat{h}_{k,x})\rho_\omega(\hat{h}_{-k,x}),\\
\scpr{\rho_\omega(\hat{h}_{k,x})[\hat{a}]}
{\rho_\omega(\hat{h}_{k,x})[\hat{b}]}_\omega\ &=\
 \scpr{[{\hat a}]} {\rho_\omega(\hat{h}_{-k,x})
\rho_\omega(\hat{h}_{k,x})[{\hat b}]}_\omega\ =\
\scpr{[{\hat a}]}{[{\hat b}]}.
\end{align}

The condition (\ref{rhohh}) is equivalent to the following condition
imposed directly on the state $\omega$
\begin{align}\label{hh-h}
\omega({\hat a}(\hat{h}_{k,y}\hat{h}_{k',y}-\hat{h}_{k+k',y}){\hat b})\ &=\ 0,\
\ \ {\rm for\ \ every}\ \ {\hat a},{\hat b}\in\kwa,
\end{align}
and every $k,k'\in \R$, $y\in M$.

We will be assuming the condition (\ref{rhohh}) in this paper. This
is equivalent to considering states on the
*-algebra
\begin{equation}\label{tildekwa}
\tilde{\kwa}\ =\ \kwa/\tilde{J},
\end{equation}
where $\tilde{J}$ is the two-sided ideal generated
by the set
\begin{equation}
\{\hat{h}_{k,y}\hat{h}_{k',y}-\hat{h}_{k+k',y}\,:\,y\in M,\ \
k,k'\in\R\}.
\end{equation}

The canonical epimorphism
$$\kwa\ \rightarrow \tilde{\kwa} $$
is used to introduce the following notation for elements of
$\tilde{\kwa}$,
\begin{align}
\hat{a}\ &\mapsto\ \tilde{a}\nonumber\\
\hat{h}_{x,k}\ &\mapsto\ \tilde{h}_{x,k}\nonumber\\
\hat{\pi}(f)\ &\mapsto\ \tilde{\pi}(f),
\end{align}
in particular $\tilde{1}$ stands for the unity in $\tilde{\kwa}$.
A technical observation extensively used below is that every
element of the quotient algebra $\tilde{\kwa}$ can be represented
by a finite linear combination of elements (\ref{prodhpi}) such
that $x_i\not= x_j$ for every pair $i\not= j$ and the hats are
replaced by the tildes. In other words, for every finitely supported
funtion  ${\bf k}:M\rightarrow \R$ denote
\begin{equation} \label{hk}
\tilde{h}_{{\bf k}}\ =\ \prod_{x\in M}\tilde{h}_{x,{\bf
k}(x)}.
\end{equation}
The algebra $\tilde{\kwa}$ is spanned by elements of the form
\begin{equation}
\tilde{h}_{\ka}, \ \ \ \  {\rm and}\ \ \ \
\tilde{h}_{\ka}\prod_{j=1}^m\tilde{\pi}(f_j),
\end{equation}
with all possible  $\ka$, $m\in\N$, and $f_1,...,f_m\in\F$.

Whereas, upon Property \ref{property}, each of the operators
$\rho_\omega(\hat{h}_{k,x})$ will be unitary, operators
$\rho_\omega(\hat{\pi}(f))$ may or may not be essentially self
adjoint, depending on a state $\omega$.

%***************************************************
\subsection{Symmetries \label{subsec:sym}}
%***************************************************

Suppose $\kwa$ is the polymer *-algebra corresponding
to a momentum variable space $\Pi_{\cal F}$ (Definition \ref{Pi})
preserved by the group Diff$(M)$ of the diffeomorphisms of $M$
(of a given class).
Every  diffeomorphism  $\varphi\in{\rm Diff}(M)$ acts
on the pair representing the classical scalar field $(\phi,\pi)$,
\begin{equation}
\phi\mapsto (\varphi^{-1})^*\phi, \ \ \ \pi\mapsto \varphi_*\pi,
\end{equation}
(notice, that $\pi$, as a signed measure, is pushed forward
by $\varphi$ rather then pulled back by its inverse).
The action passes naturally to the polymer Lie algebra and
next to the polymer *-algebra.   $\varphi$ defines a
$*$-automorphism of $\kwa$
\begin{align}
\sigma_\varphi:\kwa\ &\rightarrow \kwa\noindent\\
\sigma_\varphi(\hat{h}_{k,x})\ &= \hat{h}_{k,\varphi(x)}\label{varphi*h}\\
\sigma_\varphi(\hat{\pi}(f))&=\hat{\pi}((\varphi^{-1})^*(f)).\label{varphi*pi}
\end{align}
The automorphism $\sigma_\varphi$,
as any automorphism of $\kwa$, acts
naturally in the space of the states defined on $\kwa$. A state
$\omega$ invariant with respect to that action of the diffeomorphism group
Diff$(M)$
is called {\it diffeomorphism invariant}.

Moreover, consider an arbitrary  {\it homeomorphism} $\varphi$. In
non of the cases considered in this work the space of the smearing
functions will be preserved by all the homeomorphisms. However, it
may  be true, that a given momentum variable $\pi(f)\in\Pi_{\cal
F}$, the smearing function $f$ is mapped by $\varphi^{-1}$ into a
function $(\varphi^{-1})^*f$ such that still $(\varphi^{-1})^*f \in
{\cal F}$. The smearing functions of this property form a vector
subspace of ${\cal F}$. Denote by $\Pi_\varphi$ the corresponding
vector subspace of $\Pi$. Explicitly,
\[
\Pi_{\varphi}\ =\ \Pi_{{\cal F}\cap\varphi^*{\cal F}}
\]
It comes with the polymer *-algebra, say $\kwa_\varphi$, a subalgebra of $\kwa$.
It is obvious how to generalize the definition of the invariance of a state $\omega$
to the homeomorphisms.
The formulae (\ref{varphi*h}, \ref{varphi*pi}) define a homomorphism
\begin{equation}
\sigma_\varphi:\kwa_{\varphi}\rightarrow \kwa.
\end{equation}

\begin{df}
A state $\omega$ defined on the algebra $\kwa$ is called homeomorphism invariant
if for every homeomorphism $\varphi$
\begin{equation}
\omega(\sigma_\varphi(a))\ =\ \omega(a)
\end{equation}
for every $a\in \kwa_{\varphi}$.
\label{ho-inv}
\end{df}

Certainly, a homeomorphism  invariant state on the polymer *-algebra
$\kwa$ is also diffeomorphism invariant.

For the completeness let us explain the relation between
the diffeomorphism invariance of a state and the diffeomorphism
covariance of the corresponding representation.
A diffeomorphism invariant state $\omega$  on a polymer *-algebra
$\kwa$ defines a unitary representation $U$ of the  diffeomorphism group
Diff$(M)$ in the unitary space $\hilb_\omega$  (\ref{hilbomega})
\begin{equation}
U(\varphi)[{\hat a}]\ =\ [\varphi^*{\hat a}].
 \end{equation}
The representation $\rho_\omega$ is diffeomorphism covariant, in the
sense
\begin{equation}
U(\varphi)\rho({\hat a})U(\varphi)^{-1}\ =\ \rho(U(\varphi){\hat a})
\end{equation}
for every ${\hat a}\in \kwa$ and every $\varphi\in {\rm Diff}(M)$.

In the next section, in the case of the momentum variable space
$\Pi_{{\rm C}^{m_0}_0(M,\R)}$ we will show that a C$^{m_0}$
diffeomorphism invariant state on $\kwa$ is unique. This is the
known one used in  \cite{als}. That state is also homeomorphism
invariant.

In   Sections \ref{sec:simplicial} - \ref{sec:inv} on the other
hand, we will consider the momentum variable space $\Pi_{\cal F}$
spanned by characteristic functions of regions in $M$.  In that
case, we will additionally assume the homeomorphism invariance in
order to derive a class of new states invariant with respect to the
piecewise-analytic diffeomorphisms.

%***************************************************
\section{$\kwa$ defined by ${\rm C}^{m_0}_0(M,\R)$
smearing functions}\label{sec:smooth}
%***************************************************
In this section we choose the space ${\cal F}$ of the smearing functions
to be
\begin{equation}
{\cal F} =  {\rm C}^{m_0}_0(M,\R),
\label{F=C}
\end{equation}
that is the space of the real C$^{m_0}$-functions of compact support
and defined  on $M$, where the differentiability class  is fixed to
be either an integer $m_0\ge 0$, or  $m_0=\infty$, or C$^{(m_0)}$
stands for the semi-analytic functions according to \cite{lost}.

We  consider the polymer *-star algebra $\kwa$ corresponding to the
momentum variable space $\Pi_{\cal F}$ defined by  Definition
\ref{Pi} with the choice \eqref{F=C} of ${\cal F}$. We also identify
$\Pi_{\cal F}$ and ${\cal F}$ via the vector space isomorphism
(\ref{F2Pi}).

In this section, the group  ${\rm Diff}(M)$ considered in Section
\ref{subsec:sym} is the group ${\rm Diff}^{m_0}(M)$ of all the ${\rm
C}^{m_0}$-diffeomorphisms of $M$. It turns out that the
diffeomorphism invariance and the Property (\ref{property})
determine a state $\omega$ on $\kwa$ completely. That result, and
the proof we present below are analogous the result of \cite{lost},
but much simpler.

\begin{thr}\label{Cn}
Suppose $\kwa$ is the polymer *-algebra (Definition \ref{poly*})
defined by ${\cal F}= {\rm C}^{m_0}_0(M,\R)$. On $\kwa$, there
exists exactly one ${\rm C}^{m_0}$-diffeomorphism invariant state
$\omega_0$ such that the condition \eqref{hh-h} (equivalent to
Property \ref{property}) holds. The state is defined by
\begin{equation}\label{unique}
\omega_0({\hat a}\hat{\pi}(f))\ = 0 =\ \omega_0(\hat{h}_{k_1,x_1}\ldots
\hat{h}_{k_n,x_n}),\end{equation}
for every ${\hat a}\in \kwa$,  every momentum variable
$\pi(f)\in \Pi_{\cal F}$,
and every set of labels $\{(k_1,x_1),$ $ \ldots,(k_n,x_n)\}$
such that $x_i\not= x_j$ for every $i\not= j$,
for all $i=1,\ldots, n$ and every $n\in\N$.
\end{thr}

\begin{proof}
First, we show that if $\omega_0$ is a diffeomorphism invariant state
such Property (\ref{property}) is true, then it satisfies (\ref{unique}).
The main step is showing that
\begin{equation}\label{smoothomegapipi}
\omega_0(\gw{\hat{\pi}(f)}\hat{\pi}(f))\ =\ 0,
\end{equation}
for every $f\in{\rm C}^{m_0}_0(M,\R)$.
The trick is  to notice that

\begin{lm}\label{diffeo} For every
 $f\in {\rm C}^{m_0}_0(M,\R)$ there exists
$g\in {\rm C}^{m_0}_0(M,\R)$ and $\epsilon>0$ such that for every
$-\epsilon < \lambda< \epsilon$, there is
a $ {\rm C}^{m_0}_0(M,\R)$  diffeomorphism
$\phi_\lambda:M\rightarrow M$,
such that
\begin{equation}
g+\lambda f \ =\ \phi_\lambda^* g.
\end{equation}
\end{lm}
See \cite{lost} for the construction of $\phi_\lambda$.

 That is to say, $g+\lambda f$ is just a deformation of $g$, diffeomorphically
trivial. Then the diffeomorphism invariance implies
\begin{equation}
\begin{split}
\omega_0(\hat{\pi}(g)\hat{\pi}(g))\ &=\ \omega_0(\hat{\pi}(g+\lambda f)
\hat{\pi}(g+\lambda f))\ \\ &=\
\omega_0(\hat{\pi}(g)\hat{\pi}(g))+ 2\lambda\omega_0(\hat{\pi}(f)\hat{\pi}(g))+
\lambda^2\omega_0(\hat{\pi}(f)\hat{\pi}(f)),
\end{split}
\end{equation}
for every value $\lambda\in ]-\epsilon,\epsilon[$. Hence, the terms
proportional to $\lambda$ and $\lambda^2$ vanish.
(Recall also that $\hat{\pi}(f)^*=\hat{\pi}(f)$.) This implies the first
equality
in (\ref{unique}). Notice, that it also implies
\begin{equation}
\omega_0(\hat{\pi}(f)A)\ = \overline{\omega_0(A^*\hat{\pi}(f))}\ = 0,
\end{equation}
for every $A\in \kwa$ and every $\pi(f)\in \Pi_{\cal F}$.

Now, consider an element  ${\hat h}:=\hat{h}_{k_1,x_1}\ldots \hat{h}_{k_n,x_n}\in
\kwa$.
Certainly, for every $f\in {\rm C}^{m_0}_0(M,\R)$
\begin{equation}
0\ =\ \omega_0({\hat h}h\hat{\pi}(f)-\hat{\pi}(f)h)\ =\
\left(\sum_{i=1}^nk_if(x_i)\right)\omega_0(h).
\end{equation}
If ${\hat h}$  satisfies the assumption that  $x_i\not= x_j$ whenever $i\not= j$
 then there is
$f\in {\rm C}^{m_0}_0(M,\R)$ such that the coefficient at $\omega_0({\hat h})$ in
the equality above is not zero, hence the second equality in (\ref{unique}) is
necessarily true.
Hence, we have showed that (\ref{unique}) is a necessary condition.

Now, we show, that if $\omega_0$ satisfies (Property \ref{property})
and  (\ref{unique}),  then it is unique. Notice, that the only elements
of $\kwa$ that have not been taken into account in
(\ref{unique}) are the linear combinations of elements
$\hat{h}_{k_1,y_1}\ldots \hat{h}_{k_n,y_n}$
such that there is $i,j\in \{1,\ldots,k\}$ such that
\begin{equation}
i\not=j, \ \ \ {\rm and}\ \ \ y_i=y_j=x.
\end{equation}
However, the equations (\ref{unique}, \ref{hh-h}) and the second
equation in (\ref{state}) determine $\omega_0$ even on those
elements of $\kwa$.

Concluding, we have shown that if $\omega_0$ is a diffeomorphism invariant
state such that Property (\ref{property}) is satisfied,
then it  satisfies (\ref{unique}) and  is unique.

It is easy to construct a state $\omega_0$ by using
 (\ref{rhohh}, \ref{unique}).
The conditions  (\ref{rhohh}, \ref{unique})
define a state on  the algebra $\tilde{\kwa}$
(see \ref{tildekwa}). The state $\omega_0$ is the pullback of that state
to $\kwa$, hence it exists. Obviously, it is diffeomorphism invariant.
\end{proof}

The GNS representation $\rho_{\omega_0}$ defined by the state $\omega_0$
can be derived explicitly. The first step is to find a convenient
notation for the elements of the quotient vector space $\hilb_{\omega_0}^0$.

Every non-zero element of $[\hat{a}]\ \in\ \kwa/{J_{\omega_0}}$ can be
defined by
$\hat{a}\in\kwa$ of the following form,
 $$\hat{a}=[\prod_{i=1}^n\hat{h}_{k_i,x_i}]$$
and can be labeled by a finitely supported function
$\ka:M\rightarrow \R$,
\begin{equation}
\ka\ =\ \sum_{i=1}^nk_i\one_{\{x_i\}},
\end{equation}
where given $x\in M$, $\one_{\{x\}}$ is the characteristic
function of $x$. In particular, the identically zero function
labels the element $[\hat{1}]\ \in\ \kwa/{J_{\omega_0}}$.
Therefore we will write
\begin{align}
|\ka\rangle\ &:=\
[\prod_{x\in M}^n\hat{h}_{\ka(x),x}]
\end{align}
where $\ka:M\rightarrow \R$ has a finite support.

Therefore, $\hilb^0_{\omega_0}$ is in this case the complex vector
space $\Cyl$  of the finite formal linear combinations of elements
$|\ka\rangle$ labeled by all the functions $\ka:M\rightarrow \R$ of
{\it finite} support,
\begin{equation}\label{Cylv}
\hilb_{\omega_0}^0\ =\ \Cyl\ =\
\{\sum_{i=1}^n a_i |\ka_i\rangle\,:\, n\in\N,\,a_i\in\C,\,
\ka_i:M\rightarrow \R \ \ {\rm finitely\ \ supported}\}.
\end{equation}

Now, the GNS unitary scalar product $\scpr{\cdot}{\cdot}_{\omega_0}$
defined in $\hilb_{\omega_0}^0$ by the state $\omega_0$ is the
following unitary scalar product $(\cdot|\cdot)_{\Cyl}$:
\begin{equation} \label{Cylsc}
(|\ka_{\bf 1}\rangle\,\,|\,|\ka_{\bf 2}\rangle)_{\Cyl}\ =\
 \left\{\begin{array}{ll} 1, {\rm if}\ \ \ka_{\bf 1}\ &=\ \ka_{\bf 2}\\
0, {\rm if}\ \ \ka_{\bf 1}\ &\not=\ \ka_{\bf 2} .\\
\end{array}\right.
\end{equation}

Finally, the action  $\rho_{\omega_0}\ :\ \kwa\rightarrow\ {\rm End}(\Cyl)$
defined on the generators of $\kwa$ and the basis in $\Cyl$ can be derived:
\begin{align}
\rho_{\omega_0}(\hat{h}_{k',x})(|\ka\rangle\ &=\
[\hat{h}_{k',x}\prod_{x\in M}^n\hat{h}_{\ka(x),x}]\ =\
|\ka+k'\one_{\{x\}}\rangle\label{rhouniquea}\\
\rho(\hat{\pi}(f))(|\ka\rangle\ &=\ [\hat{\pi}(f)\prod_{x\in
M}^n\hat{h}_{\ka(x),x}]\ =\ [\,[\hat{\pi}(f)\, ,\,\prod_{x\in
M}^n\hat{h}_{\ka(x),x}]\,]\ =\nonumber\\ &= \ \sum_{x\in
M}\ka(x)f(x)|\ka\rangle.\label{rhouniqueb}
\end{align}
where the product (sum) has only
finitely many non-unit (non-zero)
factors (terms).

Some more remarks are in order. That unique state is used in LQG for
the quantum scalar field coupled with  the quantum geometry
\cite{tt6,als}. The symmetry group of the quantum geometry defined
by LQG is the group of semi-analytic diffeomorphisms. As  it was
shown in \cite{lost}, Lemma \ref{diffeo} continues to be true if the
C$^{m_0}$ differentiability class is replaced by the
semi-analyticity assumption defined therein. Therefore the theorem
is true in that case as well.

The representation $\rho_{\omega_0}$ is referred to in the
literature, as the polymer representation. For the same polymer
*-algebra $\kwa$ one may also define the standard Fock state of the
scalar field. The comparison between the polymer representation and
the Fock representation was discussed in \cite{als}.

%The state $\omega_0$ of Theorem \ref{Cn}  admits a universal extension
%to any other polymer *-algebra $\kwa'$ corresponding to any other
%momentum variable space $\Pi_{{\cal F}'}$ in the following sense:
%for {\it every} momentum variable space $\Pi_{{\cal F}'}$,
%the formulae  (\ref{hh-h},\,\ref{unique}) applied as in Theorem
%\ref{Cn} define  a state on the corresponding quantum algebra $\kwa'$.
%In particular,
%the state can be extended to the state, say $\omega_{\rm tot}$
%on  the polymer *-algebra
%$\kwa_{\rm tot}$ corresponding to the momentum variable space
%$\Pi_{\rm tot}={\rm Map}(M,\R)$. That algebra containes
%all the other polymer *-algebras.  Every other polymer *-algebra
%is embedded naturally in $\kwa_{\rm tot}$, and the restriction
%of  $\omega_{\rm tot}$ defines a state thereon.

An important subtlety of Theorem \ref{Cn} is that both: the smearing
functions  used in the definition of the basic momentum variables,
and the diffeomorphisms $M\rightarrow M$ are assumed to be of the
same class ${\rm C}^{m_0}$. The reason is that in the proof of Lemma
\ref{diffeo} (see\cite{lost}),
 the
suitable diffeomorphism is constructed from a given smearing
function. However, if we fix the differentiability class of the
diffeomorphism-symmetry group as ${\rm C}^{m_0}$, but  consider the
smearing functions of the class ${\rm C}^{m'}$ with $m'<m_0$, the
proof fails, and, in fact, the theorem is not true. (It is easy to
construct a counter example by using the counter example to the
uniqueness concerning the quantum algebra of the holonomy-flux
observable  found in \cite{lost}).

In the current work, our aim is to use that observation in
investigation of the  issue of the existence of new diffeomorphism
invariant states.

  A natural alternative to the ${\rm C}_0^{m_0}$ condition used in Theorem
\ref{Cn} is smearing the momentum variables against the
characteristic functions of subsets in $M$.  (This also one of the
choices typically used in LQG \cite{tt6}). With this choice of the
space ${\cal F}$ in  Definition \ref{Pi}, we will be able to
construct a family of new states.

%***************************************************
\section{The  CW-complex $*$-algebra}
\label{sec:simplicial}
%***************************************************************
%We will define in this section a vector space ${\cal F}_{\cal S}$
%spanned by embedded CW-complexes and its complexification $\FSC$.
%Henceforth, the manifold $M$  is endowed with a piecewise-analytic
%structure (see Appendix). The assumption ensures combinatorial
%properties similar to those in the piecewise linear case. We will
%derive homeomorphism invariant, symmetric, non-negative bilinear
%forms defined on the vector space $\F_{\Si}$. Next, we will consider
%the
%*-algebra $\exp(\odot{\FSC})$. (One can think of this as the second
%quantization of the simplexes considered as bosons). We will
%construct all the homeomorphism invariant states defined thereon.
%These results are interesting on their own. Moreover, in the next
%section they will be used in the context of the polymer *-algebra
%defined by Example 2 in Section \ref{subsec:class-Lie}.
%*************************************************
\subsection{The CW-complex vector space $\FS$}
\subsubsection{Definitions, the Euler Characteristics}
We assume in this section that the manifold $M$ is endowed with a
piecewise-analytic structure, and consider piecewise-analytic
$k$-simplexes in $M$, both notions  precisely defined in Appendix.
In particular, if $S$ is a $k$-simplex in $M$, then, depending on
$k$, $S$ is: $(0)$ a $1$-point subset of $M$ if $k=0$, $(i)$ an
`open' interval embedded in $M$ if $k=1$, $(N)$ an open ball in
$\R^{\dim M}$ embedded in $M$ with an inverse  local chard, if the
$k=\dim M$. In the last case $S$ is referred to just as a ball in
$M$. In the general case $S$ is referred to just as simplex (we skip
`piecewise-analytic').

\begin{df} Denote by
${\cal S}$ the set of all the $k$-simplexes in $M$, where $k$ ranges
from $0$ to $\dim M$.
\end{df}

Given a subset $U\in M$, by the characteristic function
$\one_U$ we mean the function
\begin{align}
\one_U:M &\rightarrow \R\\
\one_U(x)& =   \left\{\begin{array}{ll} 1,& {\rm if}\ x\in U\\
0,& {\rm if}\ x\notin U.\\
\end{array}\right.
\end{align}

\begin{df} The CW-complex vector space $\FS$ is the
real vector space spanned by the linear combinations of the
characteristic functions $\one_S$ of all the $k$-simplexes $S$ in
$M$:
$$\FS\ =\ \{\ \sum_{i=1}^n a_i\one_{S_i}\ :\ n\in \N,
a_i\in\R, S_i\in\Si\  \}. $$
\end{df}
The key identities satisfied by the characteristic functions
valid for arbitrary subsets $U_1,U_2\subset M$ are
 \begin{align}\label{idpi1}
\one_{U_1} + \one_{U_2} - \one_{U_1\cap U_2} -
\one_{U_1\cup U_2}\ &=\ 0\\
\one_{U_1\setminus U_2}\ - \one_{U_1}+ \one_{U_1\cap U_2}=0.
\end{align}
Due to them, despite of the fact, we have used only the  simplexes
to define the space $\FS$, there are many other subsets
 $U\subset M$, such that the characteristic function
$$\one_U\in \FS.$$
Indeed, this is the case whenever  $U$ is triangulable (see
Appendix).

As a warm up, and to introduce the Euler Characteristic, consider
a linear map
\begin{equation}
\chi:\FS\rightarrow\R
\end{equation}
and assume it is homeomorphism invariant in the sense, that
the number $\chi(\one_S)$ depends only on the dimension of
the simplex $S$. It is easy to see that this map is uniquely
determined by the value of $\chi(S_0)$ where $S_0$ is a 0-simplex,
meaning a one point subset of $M$. Indeed, if $S_1$ is a 1-simplex,
then we can always split it into a disjoint union of
two $1$-simplexes an one $0$-simplex, and consider
the corresponding identity between the characteristic functions
\begin{align}
S_1\ &=\ {S'}_1\cup {S'}_0\cup{S''}_1\nonumber\\
\one_{S_1}\ &=\ \one_{{S'}_1}+\one_{{S'}_0}+\one_{{S''}_1}.
\end{align}
The linearity and the homeomorphism invariance
of $\chi$ applied to the identity imply
$$ \chi(\one_{S_1})\ =\ -\chi(\one_{S'_0}).$$
Repeating exactly the same calculation for $2$-simplex up to $\dim
M$-simplex,  one can convince himself, that for every $k$-simplex
$S_k$,
\begin{equation}
\chi(\one_{S_k})\ =\ \mu_0(-1)^k,
\end{equation}
where $\mu_0$ is an arbitrary constant. The map
\begin{equation}
\label{Euler}\Si\ni S_k\ \ \mapsto\ \ \chi_{\rm E}(S_k)\ =\ (-1)^k
\in \R,
\end{equation}
where $k$ is the dimension of $S_k$, is called the Euler
Characteristic.

Thus far we have derived a necessary condition determining the
possibly existing linear map  $\chi:\FS\rightarrow\R$   up to the
constant $\mu_0$. Another  argument is needed to show, that the
linear map  $\chi$ exists at all. The existence relies on two facts.
The first one, is that for every $n$-tuple of of the simplexes
$S_1,...,S_n\in\Si$, there is a single triangulation, that is a
finite subset  $\Si'\subset\Si$ of pairwise disjoint simplexes in
$M$ such that
$$ S_i\ =\ \bigcup_{j=1}^{n_i}S'_{ij}, \ \ S'_{ij}\in \Si'$$
for every $i=1,...,n$, and $j=1,...,n_i$. The second fact is,
that the Euler Characteristic is triangulation independent,
that is for every simplex $S$ and its triangulation, that is
a decomposition
$$S\ = \bigcup_{i=1}^{n}{S_i}$$
into the union of pairwise disjoint simplexes $S_1,...,S_n$, the
identity holds
$$ \chi_{\rm E}(S)\ = \sum_{i=1}^{n}\chi_{\rm E}(S_i).$$

We hope it will not lead to any confusion if we denote by
$$ \chi_{\rm E}:\FS\rightarrow\FS $$
the linear extension of the Euler Characteristic, that is the map
(\ref{Euler}) normalized by the condition
$$ \mu_0\ =\ 1,$$
for every $0$-simplex in $M$.  We keep calling the extension
$\chi_{\rm E}:\FS\rightarrow\R $ the Euler Characteristic functional.

\subsubsection{Homeomorphism invariant, symmetric, bilinear forms}
On the CW-complex vector space $\FS$, consider a symmetric, bilinear
form
$$(\cdot|\cdot)\ :\  \FS\times\FS\rightarrow\R,$$
non-negative, that is such that
for every $s\in\FS$
$$(s|s)\ \ge 0.$$

We also assume, that the form  assigns to a pair of
simplexes $S_1,S_2\in \Si$ a number $(\one_{S_1}|\one_{S_2})$
depending only on the homeomorphic characteristics of the pair
$\{S_1,S_2\}$. Precisely, we assume that $(\cdot|\cdot)$ is
homeomorphism invariant in the following sense:

\begin{df} A bilinear form  $(\cdot|\cdot)\ :\ \F_{\Si}\times\F_{\Si}\rightarrow\R$
is called homeomorphism invariant if for every pair of simplexes
$S_1,S_2\in \Si$ and every homeomorphism $\phi:M\rightarrow M$ such
that $\phi(S_1), \phi(S_2)\in \Si$, the form satisfies
\begin{equation}
(\one_{\phi(S_1)}|\one_{\phi(S_2)})\ =\
(\one_{S_1}|\one_{S_2}).
\end{equation}
\end{df}

In this section we will find all the homeomorphism invariant,
bilinear, symmetric, non-negative forms on the CW-complex vector
space $\FS$.

The key step is the following Lemma:

\begin{prop}\label{balls}
Suppose that, $\dim M >1$ and $M$ is connected.
Suppose  $(\cdot|\cdot)\ :\ \FS\times\FS\rightarrow\R$
is a homeomorphism invariant symmetric, bilinear, non-negative form.
Then, for any two balls $B_0,B_1\in \Si$
\begin{equation}
(\one_{B_1}-\one_{B_0}|\one_{B_1}-\one_{B_0})\ =\ 0.
\end{equation}
\end{prop}

\begin{proof}
Consider first the special case of the balls, when $B_0\subset B_1$
and the difference $B_1\setminus B_0$ is a tube\footnote{It is
useful to have in mind the identification of a ball with a
cylinder}, that is, there is a local coordinate system
$(x^1,...,x^N)$ in $M$ such that
\begin{equation}
B_1\setminus B_0\ =R\ =\
\{\ (x^1,\ldots,x^N)\ | \ r_0^2\leq (x^1)^2+(x^2)^2<r^2_1, \ |x_i|<H,
\ 2<i<N\ \}.
\label{ring}
\end{equation}
We have,
$$ \one_{B_1}-\one_{B_0}\ =\ \one_R, $$
therefore, in particular $\one_R\in\FS$.
\begin{lm}\label{lem:ring}
Suppose $R\subset M$ is a tube defined in a local coordinate
system by (\ref{ring}). Suppose $M$ and  $(\cdot|\cdot)$  satisfy
the assumptions of Proposition \ref{balls}. Then,
$$(\one_R|\one_R)\ =\ 0.$$
\end{lm}
\begin{proof}
The trick consists in cutting out the tube $R$ into $n$
disjoint, homeomorphic pieces
$$ R\ =\ R_1^{(n)}\cup\dots\cup R_n^{(n)},$$
such that each two pairs of pairs $(R,R^{(n)}_I)$ and $(R,R^{(n)}_J)$ are
homeomorphic to each other and
$$\one_{R_I^{(n)}}\in \Si$$
for every $I=1,...,n$.
Indeed, we can choose some $n\in \N$ and using the same coordinates
define
$$ R^{(n)}_I\ =\
\{\ (x^1,...,x^N)\in R\ :\ x^1 = r\cos\phi,\ x^2=r\sin \phi,\ \ r\in\R,\
\  (I-1)\frac{2\pi}{n}\le\phi<I\frac{2\pi}{n}\ \}$$
Then,
$$ \one_R\ =\ \one_{R_1^{(n)}}+\dots+
\one_{R_n^{(n)}}, \ \ {\rm and}\ \ \ \one_{R_I^{(n)}}\in \FS$$
for every $I=1,...,n$.
Now, the number in question can be written in the following form,
\begin{equation}\label{ringring}
(\one_R| \one_R)\ =
 n( \one_{R}| \one_{R_1^{(n)}})
\end{equation}
Where we have used the homeomorphism invariance of the form
$(\cdot|\cdot)$. The left hand side of the equality in independent
of $n$. The right hand side obviously depends on $n$.
However,  the factor $( \one_{R}| \one_{R_1^{(n)}})$,
again due to the homeomorphism invariance, is $n$ independent!
The only solution of the equation (\ref{ringring}) valid for
arbitrary $n$ is
$$ (\one_R|\one_R)\ =\ 0.$$
\end{proof}
In the continuation of the proof of Proposition \ref{balls}
we will use the known fact, that given in a vector space
a symmetric, bi-linear and non-negative form, the `zero norm'
vectors form a vector space, that is, in our case
$$ (s|s)\ =\ 0, \ \ {\rm and}\ \ (s'|s')=0\ \ \Rightarrow\ \
(s+as'|s+as')=0$$
for every $a\in \R$.

\begin{lm}\label{lem:balls}
Suppose $B_0$ and $B'_0$ are two disjoint balls in $M$ (that is
$B_0,B'_0\in\S$ are simplexes of the maximal dimension). Suppose
$M$ and $(\cdot|\cdot)$  satisfy the assumptions of Proposition
\ref{balls}. Then,
$$(\one_{B_0}-\one_{B'_0}|\one_{B_0}-\one_{B'_0})\ =\ 0.$$
\end{lm}
\begin{proof}
Given the balls $B_0$ and $B'_0$, there is a single ball $B_1$ such
that both $(i)$ $B_0\subset B_1$ and $R=B_1\setminus B_0$
is a tube, and $(ii)$  $B'_0\subset B_1$ and  $R'=B_1\setminus B'_0$
is a tube. (The notion of `tube' is defined by (\ref{ring})).
We can write
$$\one_{B_0}-\one_{B'_0}=
\one_{B_0} - \one_{B_1} + \one_{B_1}-\one_{B'_0} =
-\one_{R}+ \one_{R'}.$$
The application of Lemma \ref{lem:ring} concludes the proof
of Lemma \ref{lem:balls}.
\end{proof}
Finally, suppose $B_1,B_2\subset M$ are arbitrary two balls in $M$.
Let $B_3$ be a ball in $M$  disjoint from $B_1\cup B_2$.
We have
$$ \one_{B_1}-\one_{B_2}\ =\ \one_{B_1}-\one_{B_3}+\one_{B_3}-
\one_{B_2}$$
and Lemma \ref{lem:balls} applies to the first difference
as well as to the second one. Hence, Proposition \ref{balls}
follows.
\end{proof}

The conclusion from Proposition \ref{balls}, is that
that if we fix a ball $B_0$ in $M$, then for every
other ball $B$ and every $s\in \FS$,
$$ (\one_B|s)\ =\ (\one_{B_0}|s) $$
provided the assumptions of Proposition \ref{balls} are satisfied.

 It follows, that the form $(\cdot|\cdot)$ can be determined
via Proposition \ref{balls} up a normalization factor. To see
it, note first:

\begin{lm} For every simplex $S$ in $M$, there exist balls
$B_1,...,B_k$, such
that
\begin{equation}\label{SB1}
 \one_{S}\ =\ \sum_{i=1}^n m_i \one_{B_i},
\end{equation}
where $m_1,\,\dots,m_n$ are integers such that
\begin{equation}\label{SB2}
\chi_{\rm E}(S)\ =\ (-1)^{\dim M}\sum_{i=1}^nm_i\, .
\end{equation}
\end{lm}

\begin{proof}
The formula (\ref{SB2}) follows directly from the application of
the Euler Characteristic functional to (\ref{SB1}).Given a $k$-simplex $S$ 
in $M$ such that $k<{\rm dim}M$, it is easy to see that there are three $k+1$ 
simplexes $S', S"$ and $S'''$ in $M$ such that 
$${\bf 1}_{S}\ =\ {\bf 1}_{S'''}-{\bf 1}_{S"}-{\bf 1}_{S'}$$ and all the terms 
on the right hand-side are characteristic functions of $k+1$-simplexes.  Repeating this observation sufficiently many times
we get (\ref{SB1}).
\end{proof}

 Therefore, we can calculate now  for arbitrary two elements $s,s'\in\FS$
\begin{equation}
(s|s')\ =\ (\sum_{i=1}^n m_i \one_{B_i}|\sum_{i'=1}^{n'} m_{i'}
\one_{B_{i'}})\ =\ \chi_{\rm E}(s)\chi_{\rm
E}(s')(\one_{B_0}|\one_{B_0}),
\end{equation}
where we have used that
$$ ( \one_{B_i}| \one_{B_{i'}} )\ =\ ( \one_{B_0}| \one_{B_{0}} ).$$

Hence, $(\cdot|\cdot)$ is determined indeed up to a constant factor.
Due to the independence of the Euler Characteristic of
trianguliation, the formula above does define a bilinear form
$(\cdot|\cdot):\FS\times\FS\rightarrow\R$, given any value
$$\lambda\ =\ (\one_{B_0}|\one_{B_0}) $$

In this way we have proved, that:
\begin{thr}\label{product}
There exist a unique, modulo the re-scaling by an arbitrary factor
$\mu_2>0$, homeomorphism invariant, bilinear, symmetric form
$(\cdot|\cdot)$ defined on the CW-complex vector space $\FS$,
provided $\dim M>1$ and $M$ is connected. The form is defined as
follows:
\begin{equation}
(\cdot|\cdot)\ =\ \mu_2\chi_{\rm E}(\cdot)\chi_{\rm E}(\cdot),
\end{equation}
where $\chi_{\rm E}:\FS\rightarrow \R$ is the Euler Characteristic
functional .
\end{thr}

Notice, finally, that it follows from Theorem \ref{product}
that the resulting unitary space
$$ \FS/\{s\in\FS : (s|s)=0\} $$
is just one dimensional. If we denote by $[\one_{B}]$ the element of the
space $ \FS/\{s\in\FS : (s|s)=0\} $ corresponding to a ball $B$ in
$M$, then the projection map is
$$\FS\ni s\mapsto (-1)^{\dim M}\chi_{\rm E}(s)[\one_{B}].$$
Notice also, that the element $[\one_{B}]$ is unique, it does not depend
on which ball $B$ in $M$ we use to define it.
%***************************************************
\subsection{Homeomorphism invariant states on $\exp(\odot\FS^\C)$}
\label{subsec:CW-complex}
%***************************************************
Consider the complexification $\FSC$ of the (real) CW-complex vector
space $\FS$. We will denote by $\odot$ the symmetrized tensor
product, and use the following notation for an element $s\in\FSC$
\begin{equation}
s^{\odot n}\ = s\odot\dots\odot s
\end{equation}
($n$ copies of $s$ on the right hand side) and extend this
notation to the space $\FSC$ itself
\begin{equation}
(\FSC)^{\odot n}\ = \FSC\odot\dots\odot\FSC,
\end{equation}
that is $(\FSC)^{\odot n}$ is the space of the symmetric elements
of the tensor product $\FSC\otimes\dots\otimes\FSC$. Consider the
following vector space
\begin{equation}
\exp(\odot\FSC)\ :=\ \bigoplus_{n=0}^\infty (\FSC)^{\odot n}.
\end{equation}
The space has the natural commutative  *-algebra structure.
Indeed, $\exp(\odot\FS^\C)$ is by definition a  complex vector
space. The associative composition operation is the symmetrized
tensor product $\odot$, and the star operation is defined by using the
complex  conjugation $\bar{\cdot}$ in $\FS^\C$,
\begin{align}
(a\odot\dots\odot b)(c\odot\dots\odot d)\ &=\ a\odot\dots\odot
b\odot c\odot\dots\odot d\nonumber\\
(a\odot\dots\odot b)*\ &=\ \bar{b}\odot\dots\odot \bar{a}.
\end{align}

\begin{df}
The *-algebra  $\exp(\odot\FSC)$ is called the CW-complex
*-algebra.
\end{df}

As we will see this algebra is naturally isomorphic to subalgebra of
$\kwa$ generated by momentum variables \ref{mathfrakP}.

\begin{df} A state $\check{\omega}:\exp(\odot\FSC)\rightarrow\C$ is called
homeomorphism invariant if for every $k$-tuple $S_1,...,S_k\in
\Si$ of simplexes in $M$ and every homeomorphism
$\varphi:M\rightarrow M$ such that still
$\varphi(S_1),...,\varphi(S_k)\in \Si$, the following is true
$$\check{\omega}(\one_{\varphi(S_1)}\odot\dots\odot\one_{\varphi(S_k)})
= \check{\omega}(\one_{S_1}\odot\dots\odot\one_{S_k}).$$
\end{df}

In this section we derive all the homeomorphism invariant states
on $\exp(\odot\FS^\C)$.

Suppose that
$${\check{\omega}}:\exp(\odot\FS^\C)\rightarrow\C $$
is a homeomorphism invariant state. It is clear that
${\check{\omega}}$ defines a homeomorphism invariant, symmetric,
bilinear form $(\cdot|\cdot):\FS\times\FS\rightarrow\R$, namely
$$(s|s')\ =\ \check{\omega}(s\odot s') $$
(recall, that the subspace $\FS\subset\FSC$ consist of the real
elements such that $\bar{s}=s$).
 Given any two balls $B_1$ and $B_2$ in $M$, Proposition \ref{balls}
guarantees that the difference $\one_{B_1}-\one_{B_2}\in\FS$ satisfies
$$\check{\omega}(\ (\one_{B_1}-\one_{B_2})\odot(\one_{B_1}-\one_{B_2})\ )
\ =\ 0.$$
Now, it follows from the basic properties of the commutative
*-algebras, that for every $A, A'\in \exp(\odot\FSC)$,
\begin{equation}
\label{AA'} \check{\omega}(\ A\odot(\one_{B_1}-\one_{B_2})\odot A'\ )\ =\
0.
\end{equation}
Using this observation, we conclude that the state $\check{\omega}$ is
completely characterized by the following sequence of the numbers
\begin{equation}
\mu_n\ :=\ \check{\omega}(\ (\one_{B_0})^{\odot n}\ ), \ \ \ \ \ n\in\N.
\end{equation}
where $B_0\in\Si$ is an arbitrarily fixed ball in $M$. Indeed,
given any finite set of balls $B_1,...,B_n$ in $M$ the identity
(\ref{AA'}) applied $n$  times implies
$$ \check{\omega}(\one_{B_1}\odot\dots\odot\one_{B_n})\ =\
\check{\omega}((\one_{B_0})^{\odot n}).$$
Furthermore, due to (\ref{SB1},\ref{SB2}), the subset of all the
elements of  $\exp(\odot\FSC)$  of the form
$\one_{B_1}\odot\dots\odot\one_{B_n}$ where $B_k$ are balls in
$M$, spans the vector space $\exp(\odot\FSC)$.

%, given arbitrary elements $s_1,...s_n\in\FS$, and using the
%equalities (\ref{SB1},\ref{SB2}) we find that
%\begin{equation}
%\omega(s_1\odot\dots\odot s_n)\ =\ \prod_{k=1}^n(-1)^{\dim
%M}\chi_{\rm E}(s_k)\omega((\one_B)^{\odot n}).
%\end{equation}

In other words, every homeomorphism invariant state ${\check{\omega}}$ on
$\exp(\odot\FSC)$ is unambiguously determined by its restriction
to the unital subalgebra $\C[\one_{B_0}]$ of $\exp(\odot\FSC)$
generated by an element ${\one}_{B_0}$, where $B_0$ is an arbitrarily
fixed ball in $M$. It is clear that $\C[\one_{B_0}]$ is isomorphic
to the $*$-algebra $\C[\tau]$ of complex valued polynomials of one real
variable; let us explain the notation:
$$\tau:\R\rightarrow \R,\ \ \ \tau(r)=r $$
is the identity map,
$$\C[\tau]\ =\ \{\ \sum_{j=1}^na_j\tau^{j-1} : n\in\N,\ a_j\in\C
\ \}.
$$
and the isomorphism $\C[\one_{B_0}] \rightarrow\C[\tau]$ is
defined just by
\begin{equation}
\one_{B_0}\ \mapsto\ \tau, \label{iso}.
\end{equation}

%Assume now that the state $\check{\omega}$ is also self-adjoint (see Definition \ref{sa-df}). Then the operator $\rho_{\check{\omega}}(\hat{\pi}_B)$, where $\rho_{\check{\omega}}$
%is the GNS representation of $\mathfrak{P}$ given by $\check{\omega}$, is (essentially) self-adjoint and
%\[
%\check{\omega}(\hat{\pi}_B^k)=\scpr{[\hat{1}]}{\rho_{\check{\omega}}(\hat{\pi}_B)^k[\hat{1}]}=\int_{\R} (\,(-1)^{\dim M}\zeta\,)^k dE_{[\hat{1}],[\hat{1}]}(\zeta),
%\]
%where $dE$ is the spectral measure of the operator $(-1)^{\dim M}\rho_{\check{\omega}}(\hat{\pi}_B)$.

Thus we have arrived at

\begin{lm}\label{omega->mu}
Every homeomorphism invariant state ${\check{\omega}}$ on the $*$-algebra
$\exp(\odot\FSC)$ is determined by a state $\mu$ defined on the
$*$-algebra $\C[\tau]$, according to the following formula
\begin{equation}
{\check{\omega}}(\one_{B_1}\ldots\one_{B_k})\ = \mu(\tau^k), \label{om-om}
\end{equation}
for every $k\in\N$.
\end{lm}

%***************************************************
%\subsection{Existence of homeomorphism invariant states on $\mathfrak{P}$}
%***************************************************
In Lemma \ref{omega->mu} the existence of the state $\check{\omega}$ defined on
$\exp(\odot\FSC)$ has been assumed. Now, we turn to the existence
issue itself. We will show, that:
\begin{lm}
Every state $\mu$ defined on the polynomial *-algebra $\C[\tau]$
defines a unique homeomorphism invariant state ${\check{\omega}}$ on
$\exp(\odot\FSC)$ such that (\eqref{om-om}).
\end{lm}

\begin{proof} Let us begin by constructing a $*$-homomorphism
$${\rm Eul}: \exp(\odot\FSC)\rightarrow \C[\tau]$$
such that for every ball $B'$ in $M$,
$$ \one_{B'}\mapsto \tau \in C[\tau].$$

The linear extension of the equality above to the vector space
$\FSC$, the complexification of $\FS$, is defined by using the Euler
Characteristic (\ref{Euler}) in the following way,
\begin{align}
 {\rm
Eul}\ :\ \FSC\ &\rightarrow\ \C[\tau],\nonumber\\
s\ &\mapsto\ (-1)^{\dim M} \chi_{\rm E}(s)\,\tau\label{Js}
\end{align}
Because $\FSC$ generates the commutative algebra $\exp(\odot\FSC)$
freely modulo the linear relations satisfied by its elements,
the above formula can be uniquely extended to the whole
algebra $\exp(\odot\FSC)$,
\begin{equation}\label{J}
{\rm Eul}(s_1\odot\dots\odot s_n)\ :=\ {\rm Eul}(s_1)\cdot\dots \cdot
{\rm Eul}(s_n)\in \C[\tau],
\end{equation}

Suppose $\mu$ is a state on $\C[\tau]$. The pullback
\[
{\check{\omega}}\ :=\ \mu\circ{\rm Eul}
\]
is certainly a state on $\exp(\odot\FSC)$. It is also
homeomorphism invariant. Finally, the state $\check{\omega}$
satisfies  (\eqref{om-om}).

\end{proof}

Let us summarize our observations by the following:

\begin{thr}\label{thr:symp} There is a 1-1 correspondence between the space
of the homeomorphism invariant states on the CW-complex
*-algebra $\exp(\odot\FSC)$ and the space of the states on the
polynomial algebra $\C[\tau]$. The correspondence is defined by the
pullback ${\rm Eul}^*$ with the *-homomorphism ${\rm
Eul}:\exp(\odot{\FSC})\rightarrow\C[\tau]$  (\ref{Js},\ref{J}).
\end{thr}

The careful reader noticed, that the polynomial algebra $\C[\tau]$
has emerged just as isomorphic to the unital subalgebra of the
CW-complex algebra, generated by a fixed single ball $\one_{B_0}$.
In fact, using the polynomial algebra representation has many
advantages. To begin with, an example of a state $\mu$ on the
polynomial algebra $\C[\tau]$ is defined by a probability (regular,
Borel) measure $d\mu$ on $\R$, via
\begin{equation}\label{me}
\mu(\tau^k)\ :=\ \int_\R \tau^k d\mu(\tau),
\end{equation}
for every $k\in\N$, provided all the functions $\tau^{k}$ are
integrable. Conversely, every state $\mu$ on the polynomial
*-algebra  $\C[\tau]$ can be constructed in that way.
This nontrivial fact follows from the existence of a self-adjoint
extension for the operator $\rho_\mu(\tau)$, where $\rho_\mu$ is
the GNS representation (see Section \ref{subsec:GNS}. The measure is
unique if the extension is unique).

Given a homeomorphism invariant state $\check{\omega}$ defined on
the CW-complex *-algebra $\exp(\odot\FSC)$ and the corresponding
state $\mu$ defined on the polynomial algebra $\C[\tau]$ we
construct now the corresponding GNS representation.

We start with the application of the GNS construction to the state
$\mu:\C[\tau]\rightarrow\C$. Consider the corresponding
GNS representation
$(\hilb_\mu^0,\,\scpr{\cdot}{\cdot}_\mu,\,\rho_\mu,\,\Omega_\mu)$
of Section \ref{subsec:GNS}.
In particular, the  unitary space is
\begin{equation}\label{hilb}
\hilb_{\mu}^0\ =\ \C[\tau]/{J_\mu},
\end{equation}
equipped with the unitary scalar product
\begin{equation}\label{scalar}
\scpr{[P_1]}{[P_2]}_{\mu}\ :=\
\mu(\overline{P_1}P_2).
\end{equation}

Since the state $\check{\omega}$ on the algebra $\exp(\odot\FSC)$ is the
pullback
obtained by the epimorphism ${\rm Eul}:\exp(\odot\FSC)\rightarrow\C[\tau]$,
there is the natural identification
\begin{equation}
(\hilb_{\check{\omega}}^0,\,\scpr{\cdot}{\cdot}_{\check{\omega}})\
=\ (\hilb_\mu^0,\,\scpr{\cdot}{\cdot}_\mu).
\end{equation}
And the action of the GNS representation $\rho_{\check{\omega}}$
becomes:
\begin{align}\label{rho2}
\rho_{\check{\omega}}\ :\ \exp(\odot\FSC)\ &\rightarrow\
{\rm End}(\hilb_{\mu}^0),
\noindent\\
\rho_{\check{\omega}}(A)P\ &:=\ [{\rm Eul}(A)\,P]\nonumber\
\end{align}
using the map (\ref{Js},\ref{J}).
Obviously,
\begin{equation}\label{1}
\Omega_{\check{\omega}}\ =\ \Omega_{\mu}\ =\ [1],
\end{equation}
meaning the (equivalence class of) the  $0$-order
polynomial $1$.

\begin{prop}
 $(\hilb_{\mu}^0, \scpr{\cdot}{\cdot}_\mu, \rho_{\check{\omega}},
\Omega_{\mu})$
 defined by (\ref{hilb},\ref{scalar},\ref{rho2},\ref{1})
above is (equi\-valent to) the GNS representation (\ref{GNSrep})
corresponding to the state $\check{\omega}$.
\end{prop}

%***************************************************
\section{$\kwa$ defined by the smearing characteristic
functions}\label{sec:char}
%***************************************************

In this section we combine the definitions and results of the
previous Section \ref{sec:simplicial} with the notion of the the
polymer *-algebra of Section \ref{sec:pol-alg}. We consider here the
polymer *-algebra $\kwa$  defined by the space of the smearing
functions $\F$ taken to be the CW-complex vector space $\FS$,
\begin{equation}\label{FFS}
\F\ =\ \FS.
\end{equation}
At the first sight it may be surprising, that in our definition we
admit also the smearing functions supported on measure zero lower
dimensional simplexes. However, we remember from the previous
section, that  the simplex vector space $\FS$ is spanned by the
characteristic functions of the open balls in $M$. Therefore the
emergence of the measure zero supports is a result of the linear
structure. Still, a state on the corresponding polymer *-algebra
$\kwa$ might just ignore (in the suitable sense) the quantum
momentum variables corresponding to those measure zero sets. We are
not making that assumption, though.

We will construct now all the homeomorphism invariant (in the sense
of Definition 2.10) states on $\kwa$ which satisfy Property
(\ref{property}). In the next sections, we will study the
corresponding GNS representations, their equivalence classes, and
the irreducibility.

The momentum variable space $\Pi_{\FS}$ (see Definition \ref{Pi})
corresponding to that space of the smearing functions  will be
denoted by
$$ \Pi\ :=\  \Pi_{\FS}.$$
Since the momentum variables are labeled by the smearing functions,
and the smearing functions are labeled in this section by (suitable)
subsets in $M$, we will often use the following notation for the
momentum variables assigned to a $k$-simplex $S$ in $M$ (or more
generally, a triangulable subset $S\subset M$),
\begin{equation}
\pi_S\ :=\ \pi(\one_S),\ \ \ \ \ \ \hat{\pi}_S\ :=\ \hat{\pi}(\one_S)
\end{equation}

\begin{df}\label{mathfrakP}
The subalgebra $\mathfrak{P}$ of the polymer *-algebra $\kwa$
generated by all the quantum momentum variables $\hat{\pi}(s)$ where
$s$ ranges the whole space $\FS$ of the smearing functions is called
the quantum momentum algebra.
\end{df}

The quantum momentum algebra $\mathfrak{P}$ is naturally isomorphic
to the CW-complex algebra $\exp(\odot\FSC)$,
 \begin{align}
\exp(\odot\FSC)\ &\rightarrow \mathfrak{P}\label{FmathfrakP}\\
\one_{S_1}\odot\dots\odot \one_{S_n}\ &\mapsto \hat{\pi}_{S_1}\dots
\hat{\pi}_{S_n},
\end{align}
where the assignment defined for all the $n$-tuples of $k$-simplexes
$S_i$, $i=1,...,n,\,\in\N$, determines the isomorphism.

Now, every homeomorphism invariant state $\omega:\kwa\rightarrow \C$
defines by the restriction a homeomorphism invariant state on
$\mathfrak{P}$. Therefore, due to Proposition \ref{balls}, for every
two balls $B_1, B_2$ in $M$,
\begin{equation}
\omega(\ (\hat{\pi}_{B_1}-\hat{\pi}_{B_2})^2\ )\ =\ 0 ,
\end{equation}
and since $(\hat{\pi}_B)^*=\hat{\pi}_B$ for every ball $B$, the equality
implies that for every $\hat{a}\in\kwa$,
\begin{equation}
\omega(\hat{a}(\hat{\pi}_{B_1}-\hat{\pi}_{B_2}))\ =\ 0\ =\
\omega((\hat{\pi}_{B_1}-\hat{\pi}_{B_2})\hat{a}).
\end{equation}
This property leads us to the following observation:

\begin{lm}\label{lm-states}
Every homeomorphism invariant state $\omega$ on $\kwa$
%which satisfies
%Property \ref{property}
vanishes on the following elements of $\kwa$
\begin{equation}\label{vanishing}
\omega(\hat{h}_{k_1,x_1})\ =\
\omega(\prod_{j=1}^n\hat{h}_{k_j,x_j})\ =\
\omega(\prod_{j=1}^n\hat{h}_{k_j,x_j}\prod_{l=1}^m
\hat{\pi}(f_l))\ =\ 0,
\end{equation}
where all $k_j\not= 0$, the points $x_j\in M$ are pairwise
different and $n>0$.
\end{lm}

\begin{proof} Consider
\begin{equation}\hat{a}\ =\
\prod_{j=1}^n\hat{h}_{k_j,x_j}\prod_{l=1}^m \hat{\pi}(f_l)
\end{equation}
such that all $k_j\not= 0$,  the points $x_j\in M$ are pairwise
different and $n>0$. It is possible to find two balls $B_1$ and $B_2$
in $M$,
such that $B_2$ does not contain any of the points $x_j$,
$j=1,...,n$, whereas $B_1$ contains exactly one of the points, say
$x_1$. Then,
\[
[\hat{\pi}_{B_2}-\hat{\pi}_{B_1},\hat{a}]=k_1\hat{a}.
\]
Acting with $\omega$ on the both sides of the above equation and
taking advantage of the fact that $\hat{\pi}_{B_2}-
\hat{\pi}_{B_1}$ is self-adjoint element of $\kwa$ we get
\begin{equation}
0\ =\ \omega(\ (\hat{\pi}_{B_2}-\hat{\pi}_{B_1})\hat{a}\ )-
\omega(\ \hat{a}(\hat{\pi}_{B_2}-\hat{\pi}_{B_1})\ )=k_1\omega(\hat{a}).
\label{oo-to}
\end{equation}
\end{proof}

But,  the action of every state $\omega:\kwa\rightarrow\C$
which satisfies the Property (\ref{property}) is determined
by the action on the elements used in Lemma \ref{lm-states}
and the restriction of $\omega$ to the quantum momentum variable
subalgebra $\mathfrak{P}$. Therefore, the following is true:
\begin{lm}\label{det}
Every homeomorphism invariant state $\omega$ on $\kwa$ which
satisfies Property \eqref{property} is determined by its restriction
to the quantum momentum subalgebra $\mathfrak{P}$.
\end{lm}

The converse is also true:

\begin{lm}\label{exist} For every state $\check{\omega}$
on the quantum momentum subalgebra $\mathfrak{P}$ there exists a
unique extension to a homeomorphism invariant state ${\omega}$ of
Property \ref{property} defined on the polymer *-algebra $\kwa$.
\end{lm}
\begin{proof}
 It is convenient in what follows, to use the quotient algebra
  $\tilde{\kwa}$ (\ref{tildekwa}), and the
canonical epimorphism  $\kwa\ \rightarrow\ \tilde{\kwa}$.

The quantum momentum subalgebra $\mathfrak{P}$ is naturally
isomorphic to its image $\tilde{\mathfrak{P}}$
 upon the map $\kwa
\rightarrow\ \tilde{\kwa}$ therefore the subalgebras and their
elements  will be identified.

 Every state $\check{\omega}$ on the
subalgebra $\mathfrak{P}$, there defines a unique $\C$-linear
extension
$$\tilde{\omega}:\tilde{\kwa}\ \rightarrow\ \C,$$
given by the formulae (\ref{vanishing}) (where all $k_j\not= 0$,
the points $x_j\in M$ are pairwise different and $n>0$)  with
$\omega$ replaced by $\tilde{\omega}$ and the hats replaced by the
tildes.

 It remains to show that the extension  is
non-negative. The extension $\tilde{\omega}$ is just the pullback
with respect to the following $\C$-linear map:
\begin{align}
P: \tilde{\kwa}\ &\rightarrow\ \mathfrak{P}\noindent\\
P(\prod_{j=1}^n\tilde{\pi}_{U_j})\ &=\
\prod_{j=1}^n\hat{\pi}_{U_j}
\nonumber\\
P(\prod_{j=1}^n\tilde{h}_{k_j,x_j}\prod_{l=1}^m \tilde{b})\ &=\ 0,
\end{align}
whenever $k_j\not= 0$, the points $x_j\in M$ are pairwise
different and $\tilde{b}\in \mathfrak{P}$. The point is, that the
map $P$ is non-negative, in the sense that for every
$\tilde{a}\in\tilde{\kwa}$ there is
$\hat{b}_1,...,\hat{b}_n\in\mathfrak{P}$, such that
\begin{equation}\label{aa}
P(\tilde{a}^*\tilde{a})\ =\ \sum_{j=1}^n \hat{b}_j^*\hat{b}_j.
\end{equation}
Indeed, every $\tilde{a}\in \tilde{\kwa}$ can be written as a
finite sum
\begin{equation} \label{tildea}
\tilde{a}\ =\ \sum_{j=1}^n\tilde{h}_{{\bf k}_j}\tilde{b}_j
\end{equation}
where ${\bf k}_j:M\rightarrow \R$, for every $j=1,...,n$ is a finitely
supported function (see (\ref{hk})).

Then $P(\tilde{a}^*\tilde{a})$ is given by (\ref{aa}).

Therefore
\begin{equation}
\tilde{\omega}\ =\ P^*\check{\omega}
\end{equation}
 is non-negative as well and defines a  state
on $\kwa$ via the pullback $\kwa\rightarrow\tilde{\kwa}$.
\end{proof}

Combining  Lemma \ref{det} and Lemma \ref{exist} with Theorem
\ref{thr:symp} leads to a complete characterization of the
homeomorphism invariant states on the polymer *-algebra $\kwa$
provided Property \ref{property} is satisfied. The states can be
labeled by states on the polynomial algebra $\C[\tau]$ (we recall,
that the following algebras are naturally identified:
$\exp(\odot\FSC)=\mathfrak{P}=\tilde{\mathfrak{P}}$ the last one
being the image of the previous one in $\tilde{\kwa}$):

\begin{thr}\label{muomega} Suppose $\kwa$ is the polymer *-algebra
defined by the space of the smearing functions (\ref{FFS}).
 There is a natural bijection between the
set of the homeomorphism invariant states on $\kwa$ which satisfy
Property \ref{property} and the set of all the states on the
polynomial algebra $\C[\tau]$. The bijection  is the pull-back
defined by  the linear map
 $\kwa\rightarrow \tilde{\kwa}\rightarrow \C[\tau]$
where $\kwa\rightarrow\tilde{\kwa}$ is the canonical projection on
the quotient (\ref{tildekwa}), whereas the map  $\tilde{\cal E}:
\tilde{\kwa}\rightarrow \C[\tau]$ is defined as follows:
\begin{align}
\tilde{{\cal E}}_{\big|\mathfrak{P}}\ &=\ {\rm Eul}\\
\tilde{{\cal E}}\tilde{h}_{{\bf k}}\ &=\ 0,
\end{align}
(where we use the notation of (\ref{Js},\ref{J},(\ref{hk})) for
every non-zero finitely supported function  $\ka$ on $M$.
\end{thr}

%***************************************************
\section{Explicit form of GNS representations of
$\kwa$}\label{explicit2}
%***************************************************
We characterize,   in this section, the GNS representations (see
 of the states described by Theorem \ref{muomega}.
We use the notation and definitions introduced in Section
\ref{subsec:GNS}.

Let $\omega$ be a homeomorphism invariant state on   $\kwa$
  whose
GNS representation $\rho_\omega$ satisfies Property \ref{property}.
Recall, that according to Theorem \ref{muomega}, the state $\omega$
is the pullback of a state $\mu$  on the polynomial algebra
$\C[\tau]$.  We will also use, that, as explained in Section
\ref{subsec:add}, $\omega$ is the pullback of a state
$\tilde{\omega}$ on the algebra $\tilde{\kwa}$ (see
(\ref{tildekwa})).

% That is, as pointed out in Section \ref{subsec:add},
%the state $\omega$ is given by the natural pullback
%of some state $\tilde{\omega}$ defined on the quotient
%algebra $\tilde{\kwa}$ (\ref{tildekwa}).

The first step is a characterization of the unitary space
$(\,\hilb_\omega^0,\,\scpr{\cdot}{\cdot}_\omega\,)$ in which the
representation is defined; that is, the quotient space
$\kwa/{J_\omega}$ endowed with the unitary scalar product
$\scpr{\cdot}{\cdot}_\omega$. For this purpose, we will use the
unitary spaces: $(\Cyl,\ (\ \cdot,|\cdot\
)_{\Cyl})$  (\ref{Cylv},\ref{Cylsc}) and 
$(\C[\tau]/J_\mu, \scpr{\cdot}{\cdot}_\mu)$
 (\ref{hilb},\ref{scalar}). We will also use the
*-algebra homomorphism ${\rm Eul}$ (\ref{J},\ref{Js})
defined also on the quantum momentum algebra
$${\rm Eul}:\mathfrak{P}\rightarrow \C[\tau],$$
via the isomorphism (\ref{FmathfrakP}).

\begin{lm}
There is a  natural vector space isomorphism
\begin{equation}
{\cal I}\ :\ \kwa/{J_\omega}\ \rightarrow \
\Cyl\otimes\,\C[\tau]/J_{\mu},
\end{equation}
unitary with respect to the unitary forms
$\scpr{\cdot}{\cdot}_\omega$ and, respectively,
$(\cdot|\cdot)\otimes\scpr{\cdot}{\cdot}_\mu$, and such that, for
every element $\hat{b}$ of the quantum momentum algebra
$\mathfrak{P}$ (see Definition \ref{mathfrakP}) and every $n$-tuple
of quantum position variables $\hat{h}_{k_1,x_1},\dots,
\hat{h}_{k_n,x_n}$, including $\hat{h}_{0,x_i}=\hat{1}$,
\begin{equation}\label{isom}
{\cal I}\ :\ [\prod_{i=1}^n\hat{h}_{k_i,x_i}\,\hat{b}]\ \mapsto\
|\sum_{i=1}^n k_i\one_{x_i}\rangle\otimes [{\rm Eul}(\hat{b})].
\end{equation}
\end{lm}

\begin{proof} We know that the vector space $\kwa/{J_\omega}$
is isomorphic to $\tilde{\kwa}/J_{\tilde{\omega}}$ (see
(\ref{subsec:add})). A general element $\tilde{a}\in \tilde{\kwa}$
can be written in the form  (\ref{tildea}). Consider the following
map,
% map (\ref{isom})
%is equivalent to
\begin{align}\label{isom2}
\tilde{\kwa}\ &\rightarrow\ \Cyl\otimes\,\C[\tau]/J_{\mu}\\
\sum_{j=1}^n\tilde{h}_{{\bf k}_j}\tilde{b}_j \ &\mapsto\
\sum_{j=1}^n |{\bf k}_j\rangle\otimes [{\rm Eul}(\tilde{b}_j)],\nonumber
\end{align}
(as in the previous section, the subalgebra
$\mathfrak{P}\subset\kwa$ is identified with its image in
$\tilde{\kwa}$). The kernel of (\ref{isom2}) is given by the left
hand side such that
\begin{equation}
\C[\tau]/{J_\mu}\ \ni\ [{\rm Eul}(\tilde{b}_j)]\ =\ 0,\ \ \ \
j=1,\dots,n.
\end{equation}
On the other hand
\begin{equation}\label{scalartensor}
\tilde{\omega}(\, ( \sum_{j=1}^n\tilde{h}_{{\bf k}_j}\tilde{b}_j)^*
\sum_{j=1}^n\tilde{h}_{{\bf k}_j}\tilde{b}_j\, )\ =\
\sum_{j=1}^n\mu( ({\rm Eul}(\tilde{b}_j))^*{\rm Eul}(\tilde{b}_j)).
\end{equation}
Therefore the kernel of (\ref{isom2}) coincides with the left ideal
$J_{\tilde{\omega}}$. Obviously, the map (\ref{isom}) is onto.
Therefore, it passes to a vector space isomorphism, the isomorphism
${\cal I}$.

The calculation (\ref{scalartensor}) shows also that ${\cal I}$ is
unitary, if $\kwa/J_\omega$ is endowed with the unitary scalar
product $\scpr{\cdot}{\cdot}_\omega$ and the space $\Cyl\otimes\,
\C[\tau]/J_\mu$ with the tensor product $(\cdot|\cdot)_{\Cyl}\otimes
\scpr{\cdot}{\cdot}_\mu$.
\end{proof}

%It will simplify the notation if we consider the isomorphism
%${\cal I}$
%just as a labeling of the elements of the GNS unitary space
%$\hilb_\omega=\kwa/{J_\omega}$, that is denote
%\begin{equation}\label{el}
%[\tilde{h}_{\ka}\otimes \tilde{b}]\ =\
%|\ka\rangle\otimes [P],
%\end{equation}
%where $\ka:M\rightarrow\R$ is a finitely supported function, and
%$[P]\in\C[\tau]/{J_\mu}$ such that
%\begin{equation}\label{elP}
%P\ =\ {\cal Eul}(\tilde{b}).
%\end{equation}

Next, we turn to the GNS representation $\rho_\omega$ of $\kwa$
defined by the state $\omega$. Our task amounts to evaluating the
action of the operators ${\cal
I}\circ\rho_\omega(\hat{h}_{k,x})\circ{\cal I}^{-1}$, and ${\cal
I}\circ\rho_\omega(\hat{\pi}(s))\circ{\cal I}^{-1}$ in
$\Cyl\otimes\hilb_{\mu}$. For every element $|\ka\rangle\otimes[P]$,
we will denote
$$ {\cal I}^{-1}|\ka\rangle\otimes[P]\ =:\
[\tilde{h}_{\ka}\tilde{b}_P].  $$

Given any quantum position variable $\hat{h}_{k,x}$, we have
\begin{align}
{\cal I}\circ\rho_\omega(\hat{h}_{k,x})\circ{\cal
I}^{-1}|\ka\rangle\otimes[P]\ &=\ {\cal
I}\circ[\tilde{h}_{k,x}\tilde{h}_{\ka}\tilde{b}_P] =\nonumber\\
{\cal I}[\tilde{h}_{\ka + k\one_{\{x\}}}\tilde{b}_P]\  &=
|\ka+k\one_{\{x\}}\rangle\otimes[P].
\end{align}
And,   given any  embedded CW-complex $s\in \FS$
\begin{align}
{\cal I}\circ\rho_\omega(\hat{\pi}(s))\circ{\cal
I}^{-1}|\ka\rangle\otimes[P]\ &=\ {\cal
I}\circ[\tilde{\pi}(s)\tilde{h}_{\ka}\tilde{b}_P] =\nonumber\\ {\cal
I}[\tilde{h}_{\ka} (\tilde{b}_{{\rm Eul}(s)\,P}+\sum_{x\in
M}\ka(x)s(x)\tilde{b}_P)]\ &= |\ka\rangle\otimes(\sum_{x\in
M}\ka(x)s(x)[P]+[{\rm Eul}(s)P]).
\end{align}
%[
%rho_\omega(\hat{\pi}(s))|\ka\rangle\otimes[P]=[\hat{h}_{\ka}\hat{\pi}_m]
%
%\,[\pi_B,\hat{h}_{\ka}]\hat{\pi}_m+\hat{h}_{\ka}\hat{\one}_B\hat{\pi}_m]=[ \%at{h}_{\ka}\, (\,(\sum_{x\in M}\one_B(x)k(x)\,)\hat{\pi}_m+\hat{\pi}_{m+1}\,%\,],
%]
%%%%%%%%%%%%%%%%%%%%%%%%%%%%%%%%%%%%%%%%%%%%%%%%%%%%%%%%%%%%%%%%%%%%%%%%%
%%%%%%%%%%%%%%%%do tego miejsca doszedlem%%%%%%%%%%%%%%%%%%%%
%poprawic wyliczenie na gorze%%%%%%%%%%%%%%%%%%%
%%%%%%%%%%%%%%%%%%%%%%%%%%%%%%%%%%%%%%%%%%%%%%%%%%
In conclusion:

\begin{thr}\label{explicit} Suppose $\kwa$ is the polymer *-algebra
defined by the space of the smearing functions (\ref{FFS}). Suppose
$\omega$ is a homeomorphism invariant state on $\kwa$ such that
Property \ref{property} is satisfied. Let
$\mu:\C[\tau]\rightarrow\C$ be the state corresponding to $\omega$
by Theorem \ref{muomega}. The GNS representation
$(\hilb_\omega^0,\scpr{\cdot}{\cdot}_\omega,\rho_\omega,
\Omega_\omega)$ corresponding to $\omega$ (see Section
\ref{subsec:GNS}) is equivalent to $(D,\scpr{\cdot}{\cdot}, {\rho},
{\Omega})$
 where (see (\ref{Cylv},\ref{Cylsc}) and (\ref{hilb}, \ref{scalar})):
 \begin{equation}
D\ =\ \Cyl\otimes\hilb^0_\mu,\label{explicitH}
 \end{equation}
equipped  with the unitary scalar product,
\begin{equation}
\scpr{\cdot}{\cdot} \ =\ (\ \cdot|\cdot\ )_{\Cyl}\otimes
\scpr{\cdot}{\cdot}_{\mu},\label{explicitpr}
\end{equation}
and with the distinguished  vector,
 \begin{equation}
 {\Omega}\ =\ |0\rangle\otimes[1],\label{explicitcyc},
 \end{equation}
and the representation ${\rho}$ is defined as follows,
\begin{align}\label{explicitrho}
\rho(\hat{h}_{k,x})|\ka\rangle\otimes[P]\ &=\
|\ka+k\one_{\{x\}}\rangle\otimes[P].\\
{\rho}(\hat{\pi}(s))|\ka\rangle\otimes [P]\ &=\
 |\ka\rangle\otimes(\sum_{x\in M}\ka(x)s(x)[P]+
 (-1)^{{\rm dim}M}\chi_{\rm E}(s)
\rho_\mu(\tau)[P]),
\end{align}
where GNS representation $(\hilb^0_\mu,\,\scpr{\cdot}{\cdot}_\mu,\,\rho_\mu,\,
\Omega_\mu)$ corresponds to the state
$\mu:\C[\tau]\rightarrow\C$.
\end{thr}
\bigskip

\begin{itemize}
\item Henceforth we will be identifying each GNS representation
$(\hilb_\omega^0,\scpr{\cdot}{\cdot}_\omega,\rho_\omega,
\Omega_\omega)$ considered in Theorem \ref{explicit} with the
corresponding representation $(D,\scpr{\cdot}{\cdot}, {\rho},
{\Omega})$ defined therein.
\end{itemize}

%***************************************************
\section{Self-adjoint extensions}
\label{s-a}
%***************************************************
\subsection{Momentum-self-adjoint representations}
We turn now, to the issue of self-adjoint extension of the GNS
representations of the  states on the polymer
*-algebra $\kwa$ considered in Section \ref{sec:char} and Section
\ref{explicit2}. Suppose $\omega:\kwa\rightarrow\C$ is a
homeomorphism invariant state considered in Theorem \ref{muomega}
and determined by a state $\mu:\C[\tau]\rightarrow\C$  on the
polynomial algebra. We will study  the corresponding representation
 $(D,\scpr{\cdot}{\cdot},\rho,
{\Omega})$ of (\ref{explicitH} -- \ref{explicitrho}) equivalent to
the GNS representation corresponding to the state $\omega$. We will
also use the Hilbert space $\hilb$ defined by the completion
\begin{equation}
{\hilb}\ :=\ \bar{D},\label{hilbomega2}
\end{equation}
as well as the GNS representation
$(\hilb^0_\mu,\,\scpr{\cdot}{\cdot}_\mu,\,\rho_\mu,\, \Omega_\mu)$
corresponding to the state $\mu:\C[\tau]\rightarrow\C$ and the
Hilbert space
\begin{equation}
\hilb_\mu\ =\ \overline{\hilb_\mu^0}.\label{hilbmu}
\end{equation}
For every $\hat{a}\in\kwa$ the operator ${\rho}(\hat{a})$ is defined
in the domain $D\subset {\hilb}$. We know that for every (real
valued) smearing function $s\in\FS$, by the GNS construction, the
operator ${\rho}(\hat{\pi}(s))$ is symmetric. That is, for every
pair of vectors $[\hat{a}],[\hat{b}]\in D$
\[
\scpr{\rho(\hat{\pi}(s))[\hat{a}]}{[\hat{b}]} \ =\
\scpr{[\hat{a}]}{\rho(\hat{\pi}(s))[\hat{b}]}.
\]
It is easy to notice, that every operator $\rho(\hat{\pi}(s))$ is
essentially self-adjoint if
\begin{equation}
\chi_{\rm E}(s)\ =\ 0.\label{chi0}
\end{equation}
Indeed, given $s\in\FS$ such that (\ref{chi0}), we have
\[
\rho(\hat{\pi}(s))\bra{k}\otimes [P]\ =\ \sum_{x\in M}
\ka(x)s(x)\,
\bra{k}\otimes [P],
\]
for every finitely supported function $\ka$ and every polynomial
$P\in\C[\tau]$. The eigenvectors $\bra{k}\otimes [P]$ span the
domain $D$ of the operator, which shows the essential
self-adjointness.

Now we will consider the case
$$\chi_{\rm E}(s)\not=0.$$

%
%we do not have is
% However, this does not mean that the (unbounded) operator
%$\rho_\omega(\hat{\pi}(s))$ is (essentially) self-adjoint. This
%observation motivates us to formulate the following definition:
%\begin{df}
%A GNS representation $\rho_\omega$ of the polymer *-algebra $\kwa$
%is momentum-self-adjoint if for every (real valued) smearing
%function $s\in\F$, the corresponding operator
%$\rho_\omega(\hat{\pi}(s))$ defined in the domain
%$\hilb_\omega^0=\kwa_\omega/J_\omega$ is essentially self-adjoint in
%the Hilbert space $\hilb_\omega$. The state $\omega$ generating a
%momentum-self-adjoint representation of $\kwa$ is also said to be
%momentum-self-adjoint. \label{sa-df}
%\end{df}
The following theorem shows that the issue of the self-adjointness
of the quantum momentum operators $\rho(\hat{\pi}(s))$ comes down to
the self-adjointness  of the corresponding GNS operators
$\rho_\mu(\tau)$ defined in the Hilbert space $\hilb_\mu$ {spanned
by polynomials}.

\begin{thr} \label{momentumsa}Let
$(D,\,\scpr{\cdot}{\cdot},\,\rho,\, {\Omega})$ be any of the
representations (\ref{explicitH}-\ref{explicitrho}) of the polymer
*-algebra $\kwa$. The following two conditions are equivalent:
\begin{enumerate}
\item For every smearing function $s\in\FS$, the operator
$\rho(\hat{\pi}(s))$ is essentially self-adjoint in the Hilbert
space ${\hilb}$ (\ref{hilbomega2});

\item the operator $\rho_\mu(\tau)$ is essentially self-adjoint in
the Hilbert space ${\hilb}_\mu$ (\ref{hilbmu}).
\end{enumerate}
\end{thr}

\begin{proof} We will show, that for every $s\in \FS$,
the images of the operators $\rho(\hat{\pi}(s))\ \pm i$ are dense in
the Hilbert space ${\hilb}$, that is, the equalities
\begin{equation}
\overline{(\rho(\hat{\pi}(s))\ \pm\ i)D}\ =\ {\hilb},
\end{equation}
are true, if and only if
\begin{equation}\label{satau}
\overline{(\rho_\mu(\tau)\ \pm\ i)\hilb_\mu^0}\ =\ \hilb_\mu.
\end{equation}
The last condition (\ref{satau}) is necessary and sufficient
 for the
essential self adjointness of the operator $\rho_\mu(\tau)$.

First step of the proof is to fix an arbitrary vector
$|\ka\rangle\in\Cyl$,  and apply the operator in question to the
subspace
\begin{equation}
\{|\ka\rangle\}\otimes\, \hilb_\mu^0\ \subset\
\Cyl\otimes\,\hilb_\mu^0\ =\ D.
\end{equation}
The image is
\begin{equation}
(\rho(\hat{\pi}(s))\ \pm i)\{|\ka\rangle\}\otimes\, \hilb_\mu^0\ =\
\{|\ka\rangle\}\otimes ((-1)^{\rm M}\chi_{\rm E}(s)\rho_\mu(\tau) +
\sum_{x\in M}k(x)s(x)\ \pm i)\hilb_\mu^0.
\end{equation}

Focus attention on  the second factor of the tensor product on the
right hand side. Suppose
\begin{equation}
\chi_{\rm E}(s)\ \not=\ 0.\label{not0}
\end{equation}
 The factor  is dense in the
Hilbert space $\hilb_\mu$, if and only if the operator
$\rho_\mu(\tau)$ is essentially self-adjoint. (On the other hand, if
we assume that $\chi_{\rm E}(s)\ =\ 0$, then the second factor is
unconditionally dense in $\hilb_\mu$.)

Now,
\begin{equation}
(\rho(\hat{\pi}(s)\ \pm\ i)\ D\ =\
\bigoplus_{|\ka\rangle\in\Cyl}\{|\ka\rangle\}\otimes ((-1)^{\rm
M}\chi_{\rm E}(s)\rho_\mu(\tau) + \sum_{x\in M}k(x)s(x)\ \pm \
i)\hilb_\mu^0.
\end{equation}
Clearly, in the case (\ref{not0}) the right hand side is dense in
${\hilb}$ if and only if the operator $\rho_\mu(\tau)$ is
essentially self adjoint. (If $\chi_{\rm E}(s)=0$, the right hand
side equals $D$ for every $\omega$ and $\mu$).
\end{proof}

\begin{df} \label{momentumsa:df} Every representation
$(D,\scpr{\cdot}{\cdot},\rho, {\Omega})$ of the polymer $*$-algebra
defined in (\ref{explicitH} - \ref{explicitrho}) which satisfies the
conditions 1 and 2 will be called momentum self adjoint.
\end{df}

Theorem \ref{momentumsa} provides a unique self-adjoint extension
for each of the  quantum momentum operators, provided certain
conditions are satisfied. Talking about representations, however, we
need a suitable extension of the  common domain of {\it all} the
operators of a given GNS representation. The appropriate  framework
can be found in \cite{schmud}. We use some elements of that
framework in the next subsection and combine with our very results.

% In the case of
%*-algebras and unbounded operators, the notion of the
%self-adjointness of all a given representation is more difficult
%then in the unbounded operator case. We will use here the
%definitions and results which can be found in \cite{}. Our goal is
%establishing a direct relation between the essential
%self-adjointness of the quantum momentum operators studied in
%Theorem \ref{} above, and the self-adjointness of the given
%representation $\rho_\omega$ according to \cite{}. In the general
%case, we will construct a natural self-adjoint extension of our
%representation.
%\begin{df}
%Representation $(\tilde{\rho},\hilb)$ is called extension of $(\rho,\hilb)$ if
%\[
%\forall a\in\kwa\ \  D(\rho(a))\subset D(\tilde{\rho}(a)),\  \tilde{\rho}(a)|_{D(\rho(a))}=\rho(a)
%\]
%\end{df}
%We will seek such an extentions which obey following property:
%\begin{df}
%Representation  $(\tilde{\rho},\hilb)$ is called self-adjoint if
%\[
%\forall a\in\kwa\ \  \rho(a)^*=\rho(a^*)^{cl}
%\]
%\label{sa}
%\end{df}
%

\subsection{Unique self-adjoined extension}\label{uniqueext}
\begin{df} A hermitian representation of a $*$-algebra $\kwa$
is a triple $(D,\scpr{\cdot}{\cdot},\rho)$ which consist of
\begin{itemize}
\item a vector space $D$ equipped with a unitary
scalar product $\scpr{\cdot}{\cdot}$,
\item an algebra homomorphism $\rho:\kwa\rightarrow{\rm End}(D)$
into the algebra of linear operators defined in $D$,
\end{itemize}
such that
\[
\scpr{\phi}{\rho(a)\psi}=\scpr{\rho(a^*)\phi}{\psi}, \ \ \ \forall
\phi,\psi\in D.
\]
Given a hermitian representation $(D,\scpr{\cdot}{\cdot},\rho)$ we
also consider  the natural Hilbert space completion
$$\hilb\ =\ \bar{D}$$
equipped with the natural extension of the unitary scalar product
denoted by the same symbol $\scpr{\cdot}{\cdot}$.
\end{df}

We call a hermitian representation
$({D}',\scpr{\cdot}{\cdot}',\rho')$ of an algebra $\kwa$ an
extension of a representation $(D,\scpr{\cdot}{\cdot},\rho)$ of
$\kwa$ if: $(i)$ $D\subset{D}'$, and $(ii)$ $\scpr{\cdot}{\cdot}'$
is an extension of $\scpr{\cdot}{\cdot}$, and $(iii)$
\begin{equation}
{\rho'(a)}|_{D}\ = \rho(a),\ \ \ \forall a\in\kwa.
\end{equation}

% A property of every
%hermitian  representation $(D, \scpr{\cdot}{\cdot},\rho)$ of any
%$*$-algebra $\kwa$, is that for every $a\in\kwa$ the closure of the
%graph of the operator $\rho(a)$ in $\bar{D}\times\bar{D}$ is again
%the graph of an operator $\rho(a)^{\rm cl}$ defined in some
%subspace $D_{a}^{\rm cl}\subset \bar{D}$, larger then $D$.  The new
%operator is called the closure of $\rho(a)$.

For every operator $\rho(a):D\rightarrow D$ defined by a hermitian
representation, the domain $D_a^*$ of the adjoint operator
$\rho(a)^*$ is an extension of $D$,
$$ D\subset D_a^*\subset \hilb.$$

The operator $\rho(a)$ naturally extends itself to the operator
adjoint to $\rho(a^*)$,
$$\rho(a^*)^*:D_a^*\rightarrow \hilb. $$
Whereas $D_a^*$ may be not preserved by the operator, it can be
easily shown that the following intersection is preserved
\begin{align}
D^*\ &:= \bigcap_{b\in \kwa}D_b^*,\label{Dstar}\\
\rho(a^*)^*(D^*)\ &\subset\ D^*,\ \ \ \forall a\in\kwa.
\end{align}

\begin{df}\cite{schmud}\label{adjointext} Given a hermitian representation
$(D,\scpr{\cdot}{\cdot},\rho)$ of a $*$-algebra $\kwa$, the triple
$({D}^*,\scpr{\cdot}{\cdot}, \rho^{*})$ is called the adjoint
extension of $(D,\scpr{\cdot}{\cdot},\rho)$, if \begin{itemize}
\item $D^*$ is defined by (\ref{Dstar}),
\item for every $a\in\kwa$, ${\rho}^*(a):D^{*}\ \rightarrow {D}^*$
is the following linear operator
\begin{equation}
{\rho}^{*}(a)\ :=\ \rho(a^*)^{*}|_{D^{*}}.
\end{equation}
\end{itemize}
\end{df}

It is not hard to check, that the adjoint extension of a hermitian
representation of a $*$-algebra $\kwa$ is a representation of $\kwa$
in the sense that
$$\rho^*(ab)\ =\ \rho^*(a)\rho^*(b).$$
In general though, we do not know if
$(D^*,\scpr{\cdot}{\cdot},\rho^*)$ is hermitian.  That property can
be ensured by some stronger conditions.

\begin{lm}\cite{schmud}\label{hermrep} Let  $(D,\scpr{\cdot}{\cdot},\rho)$ be a
hermitian representation of a $*$-algebra $\kwa$, and $(D^{
*},\scpr{\cdot}{\cdot},\rho^{*})$ be its adjoint extension. Let
$A\subset \kwa$ be the set of elements $a\in\kwa$ such that
\begin{equation}
\scpr{v}{\rho^*(a)w}\ =\ \scpr{\rho(a^*)v}{w},\ \ \ \ \forall v,w\in
D^*.
\end{equation}
 Suppose $A$ generates the algebra $\kwa$. Then
$(D^*,\scpr{\cdot}{\cdot},\rho^*)$ is a hermitian representation of
the algebra $\kwa$.
\end{lm}

We apply now the definitions formulated above and Lemma
\ref{hermrep} to the representations of the polymer $*$-algebra.
Obviously, each of the representations
$(D,\,\scpr{\cdot}{\cdot},\,\rho,\, {\Omega})$ defined by
(\ref{explicitH} - \ref{explicitrho}) of Theorem \ref{explicit}
defines just a hermitian representation $(D,\,\scpr{\cdot}{\cdot},
\,\rho)$ of the polymer $*$-algebra (by skipping the cyclic vector
$\Omega$). Consider the adjoint extension
$(D^{*},\scpr{\cdot}{\cdot}, \rho^{*})$. In that case the set
$A\subset\kwa$ introduced in Lemma \ref{hermrep} contains every
quantum position variable $\hat{h}_{k,x}$ and all those quantum
momentum variables $\hat{\pi}(s)$ which satisfy $\chi_{\rm E}(s)=0$.
Indeed, each operator $\rho(\hat{h}_{k,x})$ is unitary in $D$,
whereas every operator $\rho(\hat{\pi}(s))$ is essentially self
adjoint. But this is not sufficient to generate all of $\kwa$.
Nonetheless, in the case of a momentum self-adjoint representation
of $\kwa$ (according to Definition \ref{momentumsa:df}), we also
have $\hat{\pi}(s)\in A$ for {\it every} quantum momentum variable.
Then, the set $A$ does generate all the algebra $\kwa$. Hence, the
adjoint extension is a hermitian representation itself. We can also
somewhat simplify the definition of the domain $D^*$ in this case.
Indeed, each of the domains $D_{\hat{h}_{k,x}}^*$  is just the full
Hilbert space $\bar{D}={\hilb}$. It follows, that in (\ref{Dstar})
the elements $b=\hat{\pi}(s_1)\cdot...\cdot\hat{\pi}(s_k)$ are
sufficient to achieve the set $D^*$ (note that { $D_{\rho(a)}^*\cap
D_{\rho(b)}^*\subset D_{\rho(a+b)}^*$ and
$D_{\rho(\hat{\pi}(s_1)\cdot...\cdot\hat{\pi}(s_k))}^*=
 D_{\rho(\hat{\pi}(s_1)\cdot...\cdot\hat{\pi}(s_k)\rho(\hat{h}_{k,x}))}^*$} ),
and we arrive at the following conclusion:

\begin{cor}\label{closure} Let $(D,\,
\scpr{\cdot}{\cdot},\,{\rho},\,{\Omega})$ be one of the
representations (\ref{explicitH} -\ \ref{explicitrho}). Suppose
$(D,\, \scpr{\cdot}{\cdot},\,{\rho},\,{\Omega})$  is momentum
self-adjoint according to Definition \ref{momentumsa:df}. Then, the
adjoint extension $({D}^*,\, \scpr{\cdot}{\cdot},\, {\rho}^*)$ of
the hermitian representation $(D,\scpr{\cdot}{\cdot}, {\rho})$ of
$\kwa$ is a hermitian representation. Moreover,
\begin{equation}
D^*\ =\ \bigcap_{s_1,...,s_n\in\FS}D_{\hat{\pi}(s_1)\cdot...\cdot
\hat{\pi}(s_n)}^*.
\end{equation}
\end{cor}

We do not need a complete characterization of the spaces
$D_{\hat{\pi}(s_1)\cdot...\cdot\hat{\pi}(s_N)}^*$  used in Corollary
\ref{closure}. However it is easy to note, that all of them contain
a useful common subspace (the one in the middle below):
\begin{equation}
D\ \subset\ \Cyl\otimes D_{\tau}\ \subset {D}^*,
\end{equation}
where the subspace $D_{\tau}\subset \hilb_\mu$   (see
(\ref{hilbmu})) defined in the next paragraph, is the domain of an
essentially self-adjoint extension of the operator $\rho_\mu(\tau)$.

Now, the Hilbert space $\hilb_\mu$ is unitary isomorphic with the
space L$^2(\mathbb{R},d{\mu})$, where $d{\mu}$ is a measure\footnote{i.e. regular, Borel measure.} on
$\mathbb{R}$ such that:
\begin{itemize}
\item[a)] $\mu(\tau^n) = \int_{\R} \tau^n d{\mu}(\tau)$, \ \ \ \
$n=0,1,...,n,...$,
\item[b)] the subspace of polynomials  is dense in $L^2(\R,{d\mu})$.
\label{measure-s-a}
\end{itemize}
This measure $d\mu$ is uniquely determined by the state $\mu$
provided $\mu$ satisfies the condition 2 of Theorem
\ref{momentumsa}. The unitary isomorphism maps $\tau\in
\C[\mathbb{R}]$ into the identity function
$\tau:\mathbb{R}\rightarrow\mathbb{R}$. Henceforth, we will be
identifying the Hilbert spaces $\hilb_\mu$ and
L$^2(\mathbb{R},d\mu)$. Via that equivalence, the subspace $D_\tau$
is
\begin{equation}\label{Dtau}
D_\tau\ =\ \{f\in{\rm L}^2(\mathbb{R},d\mu)\,:\,
\int_{\mathbb{R}}{\tau'}^{2n} |f(\tau')|^2d\mu(\tau')\ <\ \infty,
\forall n\in\mathbb{N}\}.
\end{equation}

Finally:
\begin{df} \label{sadef}A hermitian representation  of a $*$-algebra $\kwa$ is
called self adjoint if it coincides with its own adjoint
extension.
\end{df}

It can be shown (see \cite{schmud}, Section 8.1) that the adjoint
extensions $(D^*, \scpr{\cdot}{\cdot},\, {\rho}^*)$ defined in
Corolary \ref{closure} are all self-adjoint hermitian
representations.

\subsection{Non-unique self-adjoined extensions}\label{non-unique}
We will generalize here, the unique extension of a momentum
self-adjoint representation introduced in the previous subsection to
a general case. Consider now an arbitrary state
$\mu:\C[\tau]\rightarrow \C$, that is do not assume the property 2
of Theorem \ref{momentumsa}. Still, there exists a measure
$d\mu^{*}$ on $\mathbb{R}$ such that the conditions $a)-b)$ hold.
The measure defines an {essentially} self-adjoint extension
$\rho_{d\mu^{*}}^{\rm cl}(\tau)$ of the operator $\rho_\mu(\tau)$.
The domain of the extension is the subspace $D_\tau$ (\ref{Dtau}).
Identifying again the Hilbert spaces $\hilb_\mu$ and
L$^2(\mathbb{R},d\mu^{*})$, the extension is defined in $D_\tau$
(\ref{Dtau}) {by},
\begin{equation}
(\rho^{\rm cl}_{d\mu^{*}}(\tau)f)(\tau')\ =\ \tau'f(\tau').
\end{equation}
therefore we will denote it just by $\hat{\tau}$,
\begin{equation}
\hat{\tau}\ :=\ \rho^{\rm cl}_{d\mu^{*}}(\tau).\label{hattau}
\end{equation}

This extension defines naturally the extension of the GNS
representation (\ref{explicitH}-\ref{explicitrho}) corresponding to
$\mu$, to a hermitian representation $(D_{d\mu^{*}},
\scpr{\cdot}{\cdot},\rho_{d\mu^{*}})$ , where $D_{d\mu^{*}}\subset
\hilb$ such that (we skip the subscript $d\mu^*$ at the scalar
product because it coincides with the scalar product in $\hilb$)
\begin{itemize}
\item $D_{d\mu^{*}}\ =\ \Cyl\otimes D_\tau$ equipped with the
scalar product of $\Cyl\otimes {\rm L}^2(\mathbb{R},d\mu^{*})$
\item {the representation is defined as follows,}
\begin{align}
{\rho}_{d\mu^{*}}(\hat{h}_{k,x})|\ka\rangle\otimes f\ &=\
|\ka+k\one_{\{x\}}\rangle\otimes f,\label{rhodmu*a}\\
{\rho}_{d\mu^*}(\hat{\pi}(s))|\ka\rangle\otimes f\ &=\
|\ka\rangle\otimes(\sum_{x\in M}\ka(x)s(x)f\ +\ (-1)^{{\rm
dim}M}\chi_{\rm E}(s)\hat{\tau}\cdot f).\label{rhodmu*b}
\end{align}
\end{itemize}
In the same way as it was explained in the previous subsection, it
can be shown that the adjoint extension  of the hermitian
representation $(D_{d\mu^{*}}, \scpr{\cdot}{\cdot},\rho_{d\mu^{*}})$
is a hermitian representation and a self-adjoint representation of
$\kwa$ in the sense of Definition \ref{sadef}.

Finally, in the case of a momentum self-adjoint representation (see
Definition \ref{momentumsa:df}), the extension introduced in the
current subsection coincides with the unique extension introduced in
Section \ref{uniqueext}.

%***************************************************
\section{Equivalence of representations}\label{sec:equiv}
%***************************************************

%Another important issue is equivalence of representations:
%\begin{df} Suppose $(D_1,\scpr{\cdot}{\cdot}_1,\rho_1)$ and $(D_2,\scpr{\cdot}{\cdot}_2,\rho_2)$ are
%two hermitian representations of $\kwa$. An unitary map
%\[
%I:\ D_1\rightarrow \ D_2,
%\]
%satisfying following condition
%\[
%I\rho_1(a)=\rho_2(a)I
%\]
%is called an intertwiner. \label{inter}
%\end{df}

\begin{df} Two representations
$(D_1,\scpr{\cdot}{\cdot}_1,\rho_1)$ and
$(D_2,\scpr{\cdot}{\cdot}_2,\rho_2)$ are called equi\-valent if there
exists a unitary space isomorphism $I:\ D_1\ \rightarrow\ D_2$
satisfying:
\begin{equation}
I^{-1}\circ \rho_2\circ I\ =\ \rho_1.\label{interdef}
\end{equation}
$I$ is called an intertwining map between the representations.
\end{df}

{We will denote by $\overline{I}$ the unique extension of the map
$I$ to the completions $\overline{D_1}$ and $\overline{D_2}$ of the
domains.}

We will study now,  the issue of the equivalence in the context of
of the momentum self-adjoint representations considered in Section
\ref{uniqueext}, as well as in the more general context of  the
representations defined in Section \ref{non-unique}. To make our
results as general as possible, we consider the (self-)adjoint
extensions of the representations. We show, that the equivalence
issue comes down to the absolute continuity of the measures on
$\mathbb{R}$ labeling the representations.

\begin{thr}\label{thr:equiv} Let $(D_1,\,\scpr{\cdot}{\cdot}_1,\,\rho_1)$ and
$(D_2,\,\scpr{\cdot}{\cdot}_2,\,\rho_2)$ be the representations of
the algebra $\kwa$ defined by: (\ref{rhodmu*a} - \ref{rhodmu*b}),
and the measures $d\mu^*=d\mu_1,\ d\mu_2$.
\begin{itemize}
\item Suppose their
adjoint extensions $(D_1{}^{*},\scpr{\cdot}{\cdot}_1,\rho_1{}^{*})$
and $(D_2{}^{*}, \scpr{\cdot}{\cdot}_2,\rho_2^*)$ are equi\-valent.
Then:
\begin{itemize}
\item The measures $d\mu_1$ and $d\mu_2$ are absolutely continuous
with respect to each other;

\item  For every unitary isomorphism
$I:\ D_1{}^{*}\rightarrow\ D_2{}^{*}$ such that
\begin{equation}\label{interthr}
I^{-1}\circ \rho^*_2\circ I\ =\ \rho^*_1
\end{equation}
there is $h\in L^2(\R,d\mu_2)$ such that
 $I$ is of the following form for every
 $\bra{k}\in\Cyl$, and every $f\in {\rm L}^2(\mathbb{R},d\mu_1)$:
\begin{equation}\label{intertwiner}
%\overline{\Cyl}\
%\overline{\otimes} L^2(\R,d\mu_1)\in
\bra{k}\otimes f\  \mapsto\ \bra{k}\otimes hf.
%\in \overline{\Cyl}\
%\overline{\otimes} L^2(\R,d\mu_2),
\end{equation}
 Moreover,
\begin{equation}\label{dmu/dmu}
 |h|^2 \ =\
\frac{d\mu_1}{d\mu_2}.
\end{equation}
\end{itemize}

\item Conversely, if the measures $d\mu_1$ and $d\mu_2$ are
absolutely continuous, the map  (\ref{intertwiner}) with
$$h\ :=\  \sqrt{\frac{d\mu_1}{d\mu_2}}$$
intertwines the representations  $(D_1,\,\scpr{}{}_1,\,\rho_1)$ and
$(D_2,\, \scpr{\cdot}{\cdot},\,\rho_2)$ (not only the adjoint
extensions).
\end{itemize}
\end{thr}

\begin{proof}
Suppose the representations
$(D_1{}^{*},\scpr{\cdot}{\cdot}_1,\rho_1{}^{*})$ and $(D_2{}^{*},
\scpr{\cdot}{\cdot}_2,\rho_2^*)$  are equivalent and
$\overline{I}:\overline{D_1}\rightarrow \overline{D_2}$ is an
intertwining map. The sketch of the proof of (\ref{intertwiner}) is
simple. Every intertwining map $I$ is determined by its action on
the cyclic vector $\bra{0}\otimes 1$. But, as we argue below, each
intertwining map $I$ has to satisfy
\begin{equation}\label{Icycl}
I\ :\ \bra{0}\otimes 1\ \mapsto\  \bra{0}\otimes h,
\end{equation}
 where $h\in {\rm L}^2(\mathbb{R},d\mu_2)$. Next,
we find that  the map $I$ determined by (\ref{Icycl}) is exactly the
one defined in (\ref{intertwiner}).  Below we go to the details. The
only technical subtlety we have to be careful about is that we are
dealing with the adjoint extension of the representation $\rho_2$.

The vector $\bra{0}\otimes 1$ is annihilated   by the following
sub-class of the momentum operators:
\begin{equation}
\rho_1(\hat{\pi}(s))\bra{0}\otimes 1\ =\ 0, \ \ \ {\rm whenever}\ \
\ \chi_{\rm E}(s)=0.
\end{equation}
This property has to be preserved by every intertwiner, therefore
\begin{equation}
\rho^*_2(\hat{\pi}(s))I(\bra{0}\otimes 1)\ =\ 0, \ \ \ {\rm
whenever}\ \ \ \chi_{\rm E}(s)=0.
\end{equation}
But every vector $\psi\in D^*_2$ such that
\begin{equation}
\rho^*_2(\hat{\pi}(s))\psi\ =\ 0, \ \ \forall s\in\FS:\chi_{\rm
E}(s)=0,\label{anihil}
\end{equation}
is necessarily of the form $\bra{0}\otimes h$ (that statement is
more obvious in the case of $\psi\in D_2$ and $\rho_2$, and it
generalizes to the adjoint extensions), hence:
\begin{equation}
I(\bra{0}\otimes 1)\ =\ \bra{0}\otimes h,\ \ \ \ h\in {\rm
L}^2(\mathbb{R},d\mu_2).
\end{equation}
Next we determine the action of $I$ on the vectors
$\bra{0}\otimes\tau^n$, $n=1,...$ . We have
\begin{equation}
I(\bra{0}\otimes\tau^n)\ =\ \rho_2^*(\hat{\pi}(s_1))
\cdot...\cdot\rho^*_2(\hat{\pi}(s_n)) \bra{0}\otimes h,
\label{taun}
\end{equation}
 with any
$s_1,...s_n\in\FS$ such that $\chi_{\rm E}(s_i)=1$, $i=1,...,n$.
Again, we observe that the right hand side of (\ref{taun}), if
denoted by $\psi$, satisfies (\ref{anihil}). Hence, for every
$s\in\FS$,  the operator $\rho^*_2(\hat{\pi}(s))$ preserves the
subspace ${\bra{0}\otimes {\rm
L}^2(\mathbb{R},d\mu_2)\,\cap\,D^*_2}$. Fix $s$ such that $\chi_{\rm
E}(s)=1$ and consider the restriction
\begin{equation}
\rho^*_2(\hat{\pi}(s)): {\bra{0}\otimes {\rm
L}^2(\mathbb{R},d\mu_2)\,\cap\,D^*_2}\ \rightarrow {\bra{0}\otimes
{\rm L}^2(\mathbb{R},d\mu_2)\,\cap\,D^*_2}.\label{restr}
\end{equation}

It follows from (\ref{rhodmu*b}) and (\ref{hattau}) that the
operator (\ref{restr}) coincides with the operator $\hat{\tau}$ (as
expected). Therefore, we have derived
\begin{equation}
I(\bra{0}\otimes\tau^n)\ =\ \bra{0}\otimes\hat{\tau}^n\cdot h\ =\
\bra{0}\otimes h\tau^n.
\end{equation}
Finally, the action of the intertwiner $I$ on elements
$\bra{k}\otimes \tau^n$ is determined in the obvious way:
$$I (\bra{\ka}\otimes \tau^n)\ =\
I(\rho_1(\prod_{x\in M}\hat{h}_{\ka(x),x}))\bra{0}\otimes \tau^n\ =\
\rho^*_2(\prod_{x\in M}\hat{h}_{\ka(x),x}))\bra{0}\otimes h\tau^n\
=\ \bra{\ka}\otimes h\tau^n.
$$
By the continuity the map extends to (\ref{intertwiner}).

The absolute  continuity of the measures $d\mu_1$ and $d\mu_2$ is obvious
as well as the Nikodym-Radon derivative (\ref{dmu/dmu}).

Conversely, given two measures $d\mu_1$ and $d\mu_2$ on $\mathbb{R}$
which are absolutely continues, one can check by inspection that the
map (\ref{intertwiner}) is unitary, maps $D_1\rightarrow D_2$ and
intertwines the representations $\rho_1$ and $\rho_2$.
\end{proof}

%***************************************************
\section{Invariant subspaces}\label{sec:inv}
%***************************************************
In this section we solve the issue of the irreducibility of the
representations of the polymer $*$-algebra $\kwa$ introduced in
(\ref{explicitH},\ref{explicitpr},\ref{explicitrho}) and in Sections
\ref{uniqueext}, \ref{non-unique}.

Consider a representation $(D,\,\scpr{\cdot}{\cdot},\,\rho)$
(\ref{explicitH},\ref{explicitpr},\ref{explicitrho}) of the algebra
$\kwa$. As before, we will also use  the corresponding Hilbert space
 $\hilb$ (\ref{hilbomega2}). It is defined by the completed tensor
 product {(we denote it by $\overline{\otimes}$ )} of
the following spaces
\begin{equation}
\overline{\Cyl},\ \ \ \ \ \ \hilb_\mu\ =\
\overline{\C[\tau]/J_{\mu}},
\end{equation}
 the completions with respect to the unitary scalar products $( \
\cdot|\cdot\ )_{\Cyl}$, and, respectively
$\scpr{\cdot}{\cdot}_{\mu}$, that is
\begin{equation}
\hilb\ =\ \overline{\Cyl}\  \overline{\otimes}
\hilb_\mu.\label{hilbomega3}
\end{equation}
We present a quite strong, general result which
shows that the issue boils down to the irreducibility of the
representations of the polynomial algebra $\C[\tau]$ corresponding
to the state $\mu:\C[\tau]\rightarrow \C$.

Recall, that each quantum position operator $\rho(\hat{h}_{k,x}):
\hilb \rightarrow \hilb$ is an unitary {operator after extension to
the whole space}. Another class of unitary operators can be defined
by using the quantum momentum operators $\rho(\hat{\pi}(s))$
corresponding to those complexes-smearing-functions $s\in \FS$ which
satisfy
\begin{equation}
\chi_{\rm E}(s)\ =\ 0.
\end{equation}
Each of those operators with the domain $D$ (\ref{explicitH}) is
essentially-self adjoint.
 Therefore, the operator $\exp \big( i \rho(\hat{\pi}(s))\big)$
{can be defined} uniquely as an unitary operator in $\hilb$. We will
assume below that the unitary operators mentioned above preserve a
common subspace in $\hilb$, and we will come to quite strong
conclusion.
\medskip

\begin{thr}\label{inv} Let $(D,\scpr{\cdot}{\cdot},\rho)$ be a
representation of the polymer $*$-algebra defined by
(\ref{explicitH}, \ref{explicitpr}, \ref{explicitrho}),
corresponding to a state $\omega:\kwa\rightarrow\C$ determined by a
state $\mu:\C\rightarrow\C$. Suppose $\tilde{\hilb}$ is a Hilbert
subspace of the Hilbert space $\hilb$ (\ref{hilbomega3}) such that
\begin{align}
\rho(\hat{h}_{k,x})(\tilde{\hilb})\ &\subset\
\tilde{\hilb}\\
\exp\big(i\,\rho(\hat{\pi}(s))\big)(\tilde{\hilb})\ &\subset\
\tilde{\hilb},
\end{align}
for every quantum position variable $\hat{h}_{k,x}$ and every
quantum momentum $\hat{\pi}(s)$ such that (\ref{chi0}). Then, there
is a Hilbert subspace $\tilde{\hilb}_\mu$ of $\hilb_\mu$ such that
\begin{equation}
\tilde{\hilb}\ =\ \overline{\Cyl}\ \overline{\otimes}
\tilde{\hilb}_\mu.
\end{equation}
 \label{irr-thr}
\end{thr}
\begin{proof}
Given $s\in\FS$, every eigenvalue of the operator
$\exp\big(i\,\rho_\omega(\hat{\pi}(s))\big)$ can be written as
\begin{equation}
\lambda_{\ka,s}\ =\ e^{i\sum_{x\in M}\ka(x)s(x)},
\end{equation}
where $\ka$ is a finitely supported function on $M$, and for every
$\ka$ the number $\lambda_{\ka,s}$ is an eigenvalue of the operator
$\exp\big(i\,\rho_\omega(\hat{\pi}(s))\big)$. To each of the
eigenvalues $\lambda_{\ka,s}$, there is assigned  the corresponding
subspace $\hilb_{\lambda_{\ka,s}}$ of the eigenvectors. We have,
\begin{equation}
\bra{\ka'}\otimes \hilb_\mu\ \subset\ \hilb_{\lambda_{\ka,s}},
\end{equation}
for every finitely supported function $\ka'$ such that
$$\sum_{x\in M}\ka'(x)s(x)\ =\ \sum_{x\in M}\ka(x)s(x). $$
However, if we fix a finitely supported function $\ka$ and consider
the common part of all the sets $\hilb_{\lambda_{\ka,s}}$, then it
is easy to show that
\begin{equation}\label{Hlambda}
\bigcap_{s\in\FS:\ \chi_{\rm E}(s)=0}\hilb_{\lambda_{\ka,s}}\ = \
\bra{\ka}\otimes \hilb_\mu.
\end{equation}

Let us turn now, to the preserved subspace $\tilde{\hilb}$.
Denote by
$$ P_{\tilde{\hilb}}\ :\ \hilb\ \rightarrow\
\tilde{\hilb} $$
the orthogonal projection onto $\tilde{\hilb}$. For every fixed
$s\in \FS$ such that (\ref{chi0}), and a finitely supported function
$\ka$, due to the assumption that $\tilde{\hilb}$ is preserved by
the unitary operator $\exp\big(i\,\rho_\omega(\hat{\pi}(s))\big)$
and by the general properties of the unitary operators, it follows
that
\begin{equation}
P_{\tilde{\hilb}}(\hilb_{\lambda_{\ka,s}})\ \subset\
\hilb_{\lambda_{\ka,s}}.
\end{equation}
Therefore
\begin{equation}
P_{\tilde{\hilb}}(\bra{\ka}\otimes \hilb_\mu)\ \ =
\bigcap_{s}P_{\tilde{\hilb}}(\hilb_{\lambda_{\ka,s}})\
\subset\ \bra{\ka}\otimes \hilb_\mu,
\end{equation}
where the range of $s$ on the right hand side is the same as in
(\ref{Hlambda}). Hence, there is a Hilbert subspace
$\tilde{\hilb}_{\ka,\mu}\subset \hilb_\mu$, such that
\begin{equation}
P_{\tilde{\hilb}}(\bra{\ka}\otimes \hilb_\mu)\ =\
\bra{\ka}\otimes \tilde{\hilb}_{\ka,\mu}.
\end{equation}
The relevance of this space consists in the following orthogonal
decomposition of the invariant space $\tilde{\hilb}_\omega$,
\begin{equation}
\tilde{\hilb}\ =\
P_{\tilde{\hilb}}(\overline{\bigoplus_{\ka} \bra{\ka}\otimes
\tilde{\hilb}_{\mu}})\ =\ \overline{\bigoplus_{\ka} \bra{\ka}\otimes
\tilde{\hilb}_{\ka,\mu}},
\end{equation}
where $\ka:M\rightarrow\R$ ranges the set of all the finitely
supported functions.
%It is also true that
%\begin{equation}
%\bra{\ka}\otimes \tilde{\hilb}_{\ka,\mu}\ = \tilde{\hilb}\cap
%\hilb_{\ka}
%\end{equation}

Next we will show, that in fact the subspaces
$\tilde{\hilb}_{\ka,\mu}$ are necessarily all the same, independent
of $\ka$. To see that, we introduce  for every finitely supported
function $\ka:M\rightarrow\R$ the following unitary operator,
\begin{equation}
U_{\ka}\ = \prod_{x\in M}\rho(\hat{h}_{\ka(x),x}).
\end{equation}
By the definition of the operators $U_{\ka}$, and
$\rho(\hat{h}_{k,x})$,
\begin{equation}
U_{\ka}(\bra{\ka'}\otimes \hilb_\mu')\ =\
\bra{\ka+\ka'}\otimes\hilb'_\mu,
\end{equation}
for every two finitely supported functions $\ka$ and $\ka'$ defined
on $M$, and every Hilbert subspace $\hilb_\mu'\subset\hilb_\mu$. On
the other hand, the invariant space $\tilde{H}$ is in particular
both $U_{\ka}$ and $(U_{\ka})^{-1}=U_{-\ka}$ invariant, hence
\begin{equation}
U_{\ka}(\tilde{\hilb})\ =\ \tilde{\hilb}.
\end{equation}
In conclusion,
\begin{align}
\bra{\ka}\otimes\tilde{\hilb}_{\ka,\mu}\ &=\ U_{\ka}(\bra{0}\otimes
\tilde{\hilb}_{\ka,\mu})\ =\\
U_{\ka}(\bra{0}\otimes\hilb_\mu\,\cap\,\tilde{\hilb})\ &=
U_{\ka}(\bra{0}\otimes\tilde{\hilb}_{\bf{0},\mu})\ =\\
&=\ \bra{\ka}\otimes\tilde{\hilb}_{\bf{0},\mu}.
\end{align}
Therefore, the subspace pointed out in the conclusion of this
theorem is
$$  \tilde{\hilb}_{\mu}\ =\ \tilde{\hilb}_{\bf{0},\ka}. $$
\end{proof}

To complete the conclusion of Theorem \ref{inv}, consider a
hermitian representation $(D,\,\scpr{\cdot}{\cdot},\,\rho)$ of
$\kwa$. Suppose it is  either defined by a momentum self-adjoint GNS
representation and by formulae
(\ref{explicitH},\ref{explicitpr},\ref{explicitrho}), or it is one
of the representations (\ref{rhodmu*a},\ref{rhodmu*b}). It any case,
each of the operators $\rho((\hat{\pi}(s)))$, $s\in\FS$, is
essentially self adjoint, hence it defines a unitary operator
$\exp(i\,\rho(\hat{\pi}(s)))$. Suppose the subspace
$$\tilde{\hilb}\ =\ \overline{\Cyl}\bar{\otimes}\tilde{\hilb}_\mu$$ of
Theorem \ref{inv} is preserved by each of the operators
$\exp(i\,\hat{\pi}(s)))$. Then the subspace
$$\tilde{\hilb}_\mu\subset {\rm L}^2(\mathbb{R},d\mu^*)$$
is necessarily preserved by the operator $\exp(i\lambda\hat{\tau})$,
$\lambda\in\R$ (see \ref{hattau}). On the other hand, generically,
it is not hard to find such a subspace $\tilde{\hilb}_\mu$. To see
an example, suppose there is a $d\mu^*$-measurable, proper subset
$V\subset\mathbb{R}$. Define a Hilbert subspace
$\tilde{\hilb}_{\mu}\subset {\rm L}^2(\mathbb{R},d\mu^*)$ to be
spanned by (classes of) all the square integrable functions of
supports contained in $V$. The Hilbert space
$\tilde{\hilb}=\overline{\ Cyl}\
\overline{\otimes}\tilde{\hilb}_{\mu}$ has the non-trivial { (dense
in $\tilde{\hilb}$) } intersection with the domain $D^*$ of the
adjoint extension
\begin{equation}
\tilde{D}\ =\ D^*\cap \tilde{\hilb}.
\end{equation}
The subspace $\tilde{D}$ is invariant with respect to  the
representation $\rho^*$. Hence, the restriction of $\rho^*$ to
$\tilde{D}$ is a new hermitian representation of the algebra $\kwa$.

That representation is equivalent to the adjoint extension
$(D^*_{d\mu'}, \scpr{\cdot}{\cdot}',\rho^*_{d\mu'})$ of the
representation $(D_{\mu'},\scpr{\cdot}{\cdot}_{\mu'}, \rho_{\mu'})$
defined by (\ref{rhodmu*a},\ref{rhodmu*b}) and a measure
$$d\tilde{\mu}'\ =\ \one_Vd\mu^*.$$

The only case in which there is no invariant subspace
$\tilde{\hilb}$ in Theorem \ref{inv}, is the state
$\mu:\C\rightarrow\C$ defined by a measure $d\mu=\delta_{r_0}$
supported at a single point  $r_0\in\mathbb{R}$.

\section{Discussion}\label{sec:disc}
\subsection{Summary of the results}
The polymer $*$-algebra $\kwa$ (Definition \ref{poly*}) contains the
momentum subalgebra generated by  the smeared momentum variables
(Definition \ref{pi}). The study of diffeomorphism invariant states
on the momentum subalgebra is crucial for  characterization of
possible diffeomorphism invariant states on $\kwa$.

 In the C$^{(n)}$, compactly supported smearing functions case
 (Sec. \ref{sec:smooth}), there is no non-trivial
C$^{(n)}$-diffeomorphism invariant state on the momentum subalgebra.
This is the meaning  of the identity (\ref{smoothomegapipi}) derived
from the invariance. This fact and the assumption Property
\ref{property} determine the unique state on $\kwa$ (\ref{unique}).
The corresponding representation
(\ref{rhouniquea},\,\ref{rhouniqueb}) of $\kwa$ is the one used in
LQG in the quantization of the scalar field. The uniqueness  was not
a surprise, and the argument used in the proof was similar to that
of \cite{lost}.

The remaining part of the work was focused on the case of the
momentum variables $\pi(s)$ defined  by smearing against  the
characteristic functions of regions in $M$. The result is the
construction of new states. Now, the momentum subalgebra -identified
with the CW-complex algebra $\exp(\odot\FS^\C)$ in this case (Sec.
\ref{subsec:CW-complex})-  does admit non-trivial, diffeomorphism
invariant states. In order to construct explicit examples, we
assumed a greater symmetry, namely the homeomorphism invariance (see
Sec. \ref{subsec:sym}). We succeeded in deriving  all the states.
The complete class is characterized in Theorem \ref{thr:symp}.

Again, each of the states defined on the momentum subalgebra
determines a state on the polymer $*$-algebra $\kwa$ upon the
assumption Property \ref{property}. This leads to the derivation of
all the homeomorphism invariant states defined on the polymer
$*$-algebra $\kwa$ which satisfy Property \ref{property} (Theorem
\ref{muomega}). The states are labeled by all the states
$\mu:\C[\tau]\rightarrow\C$ in a 1-1 manner, where $\C[\tau]$ is the
$*$-algebra of the polynomial functions on $\R$, and $\tau={\rm
id}:\R\rightarrow\R$.

The resulting GNS representations of $\kwa$ corresponding to the
derived states, are provided in explicit form in Theorem
\ref{explicit}. The properties of the representations are analyzed
in Sec. \ref{s-a} - \ref{sec:inv}.

Given any of the  representations $\rho$ of $\kwa$, the momentum
operators $\rho(\hat{\pi}(s)$ may or may not be essentially
self-adjoint. The necessary and sufficient conditions for that
momentum self-adjoitness  are formulated in terms of the labeling
state $\mu:\C[\tau]\rightarrow\C$ in Theorem \ref{momentumsa}.

Whereas the self-adjoint extension of each of the essentially
self-adjoint momentum operators is quite well understood, the
corresponding extension of the entire representation $\rho$ requires
the special care. The suitable framework was derived by Schmudgen
\cite{schmud}. We show that for each of our momentum self-adjoint
representations $\rho$ of $\kwa$, the adjoint extension $\rho^*$
(Definition \ref{adjointext}) is a self-adjoint (Definition
\ref{sadef}) hermitian representation of $\kwa$. The remaining
representations we found are also extended to homeomorphism
invariant, self-adjoint according to \cite{schmud}, hermitian
representations of $\kwa$. Each  resulting self-adjoint
representation $\rho_{d\mu}^*$ of $\kwa$ is determined by a measure
$d\mu$ on $\R$, provided the algebra of the polynomials is dense in
L$^2(\R,d\mu)$. This is the outcome of Sec. \ref{s-a}.

Two self-adjoint representations $\rho_{d\mu_1}^*$ and
$\rho_{d\mu_2}^*$ constructed in Sec. \ref{s-a} are equivalent if
and only if the measures $d\mu_1$ and $d\mu_2$ are absolutely
continues with respect to each other (Theorem \ref{thr:equiv}). The
intertwining Hilbert space isomorphism is found  as well.

The issue of the reducibility of the representations considered
above is solved in Sec. \ref{sec:inv}.  A general characterization
of invariant subspaces is given (Theorem \ref{inv}). The final
conclusion is:

\begin{cor}
Suppose a representation $(\rho, D, \scpr{\cdot}{\cdot})$ of the
polymer $*$-algebra $\kwa$ is either
\begin{itemize}
\item the representation defined by (\ref{explicitH}, \ref{explicitpr}, \ref{explicitrho})
and a state $\mu:\C[\tau]\rightarrow\C$, and assume it is momentum
self-adjoint;

\item the representations (\ref{rhodmu*a},\ref{rhodmu*b})
 defined by a measure $d\mu^*$ on $\R$.
\end{itemize}
Then, there is no proper subspace $\tilde{\hilb}\subset \bar{D}$
preserved by the unitary extensions of all the operators
$\exp(i\rho(\hat{\pi}(s))$ and $\rho(\hat{h}_{k,x})$, unless the
state $\mu$ (the measure $d\mu^*$) is  the delta measure
$\delta_{\tau_0}$ supported at $\tau_0\in\R$. The definition of the
representation reads:
\begin{align}
(D,\scpr{\cdot}{\cdot})\ &=\ (\Cyl,(\cdot|\cdot)_{\Cyl})\nonumber\\
\rho(\hat{h}_{k,x})|\ka\rangle\ &=\
|\ka+k\one_{\{x\}}\rangle,\nonumber\\
{\rho}(\hat{\pi}(s))|\ka\rangle\ &=\
 (\sum_{x\in M}\ka(x)s(x)+
 \tau_0(-1)^{{\rm dim}M}\chi_{\rm E}(s))|\ka\rangle.
\end{align}
If $\tau'_0\not=\tau_0$, then the representation $\rho'$
corresponding to $\tau'_0$ is inequivalent $\rho$.
\end{cor}

\subsection{Relevance for the holonomy-flux algebra}
The polymer $*$-algebra considered in this work is used in Loop
Quantum Gravity for the quantum scalar field coupled with the
quantum geometry. The quantum geometry itself is described in terms
of the holonomy-flux $*$-algebra \cite{lost}. The holonomy-flux
$*$-algebra has a similar structure to the polymer $*$-algebra, and
one could say that the latter is a simplified version of the former
one. In particular, the quantum flux variable $\hat{P}(f)$ -the
counter part of the quantum momentum operators- is defined by a
smearing function $f:M\rightarrow {\rm su}(2)$ of the support
contained in an oriented 2-simplex $s$ in a $3$-dimensional
piecewise-analytic manifold $M$, and taking values in the Lie
algebra su(2). In particular, $f$ is often assumed to be of the form
\begin{equation}
f\ =\ \one_s \xi, \ \ \ \ \ \ \xi\in {\rm su}(2).
\end{equation}
The corresponding momentum can be denoted by $\hat{P}_{s,\xi}$.
Naturally, the question arises, if the class of the topological
states defined on the simplex algebra in Section
\ref{sec:simplicial} of the current paper can be used to define a
class of new diffeomnorphism invariant states on the holonomy-flux
algebra. Indeed, by exactly the same argument as in Section
\ref{sec:simplicial} we prove, that for every topologically
invariant state $\omega$
\begin{equation}\label{P-P}
 \omega((\hat{P}_{s,\xi} -
\hat{P}_{s',\xi})^*(\hat{P}_{s,\xi} - \hat{P}_{s',\xi}))\ =\ 0,
\end{equation}
for two arbitrary piecewise-analytic $2$-simplexes $s$ and $s'$ in
$M$. However, in this case a diffeomorphism which flips the
orientation of $s$ into the opposite  one is, on the one hand, a
symmetry of $\omega$ but on the other hand maps
\begin{equation}\label{-P}
\hat{P}(s,\xi)\ \mapsto\ -\hat{P}_{s,\xi}.
\end{equation}
It follows from (\ref{P-P},\ref{-P}) that
$$\omega( \hat{P}^*_{s,\xi}\hat{P}_{s,\xi})\ =\ 0.$$
Eventually, the only state is the one used in LQG. However, if we
relaxed the topological invariance assumption, in favor of proper
diffeomorphism invariance,  perhaps new states could be found on the
holonomy-flux algebra defined by the characteristic functions. On
the other hand we have the uniqueness result \cite{lost} valid in
the case of the compactly supported, C$^{(n)}$-smearing functions
$f$ (analogous to Section \ref{sec:smooth}).

\appendix

\section{Piecewise-analytic simplexes and manifolds}
In the main part of this work we are using  piecewise-analytic
simplexes embedded in a manifold $M$ endowed with an (appropriately
defined) piecewise-analytic  structure. Those notions are defined
below. The property of the piecewise-analytic simplexes crucial in
this paper is Property \ref{trian} below.

\begin{df}\label{semi-analyticset}\cite{semianalytic}
A  subset $X\subset \mathbb{R}^N$ is called semi-analytic if it has
the following property: for every $x\in \bar{X}$ (the completion of
$X$) there is a neighborhood $U\ni x$ in $\mathbb{R}^N$, and
analytic functions $f_{ij}:U\rightarrow \mathbb{R}$, $i=1,...,n,\
j=1,...,n_i$ such that
\begin{align} U\cap X \ &=\ \bigcup_{i=1}^nX_i\nonumber\\
X_i\ &=\ \{x\in U\ :\ f_{ij}(x)\leq 0, j=1,...,n_i\}.
\end{align}
\end{df}

Obviously,  every analytic diffeomorphism maps semi-analytic sets
into semi-analytic sets. However, the class of the diffeomorphisms
of that property is larger. Therefore, we define:

\begin{df} A C$^{(n)}$-diffeomorphism $\phi:U\rightarrow U'$, where $U$,
and $U'$
are open subsets in $\mathbb{R}^N$, $0\le n < \infty$, is called
piecewise-analytic if for every  semi-analytic subset $X$ of
$\mathbb{R}^N$, the subsets $\phi(X\cap U)$ and $\phi^{-1}(X\cap
U')$ are semi-analytic.
\end{df}

A sub-family of piecewise-analytic diffeomorphisms was introduced in
\cite{lost} and called semi-analytic diffeomorphisms. Many examples
were constructed. In particular, it was shown, that for every point
$x\in \mathbb{R}^N$, and every neighborhood $U$ of $x$, there is a
semi-analytic diffeomorphism which moves $x$ but is the identity map
on $\mathbb{R}^N\setminus U$. In this sense the structure we are
fixing in $\mathbb{R}^N$ to consider the semi-analytic sets admits
local degrees of freedom.

Given a manifold $M$, a structure compatible with the semi-analytic
sets can be defined as follows. Let $M$ be a differentiable (or
topological) manifold of the differentiability class C$^{(n)}$,
$0<n<\infty$ (respectively, $n=0$). Suppose the maximal atlas admits
a sub-atlas    $((U_I, \chi_I))_{I\in {\cal I}}$ labeled by some set
${\cal I}$, such that
 for every two chard    $\chi_{I}$ and $\chi_{J}$, the map
$$\chi_{J}\circ\chi_{I}{}^{-1}: \chi_{I}(U_{I}\cap U_{J})\rightarrow
\chi_J(U_{I}\cap U_{J})$$
is a piecewise-analytic C$^{(n)}$-diffeomorphism. We call the family
$((\chi_I, U_I))_{I\in\cal{I}}$ a piecewise analytic atlas on $M$
and extend to a maximal piecewise analytic atlas. A chard $(\chi,U)$
belonging to the maximal piecewise-analytic atlas is called a
piecewise-analytic chard.

\begin{df}
A manifold endowed with a piecewise analytic atlas is called
piecewise analytic.
\end{df}
Obviously, every analytic manifold is  piecewise analytic. A
piecewise analytic diffeomorphism $M\rightarrow M$ of a piecewise
analytic manifold $M$ is a diffeomorphism such that itself and its
inverse  preserve the maximal piecewise analytic atlas.

Now we turn to semi-analytic simplexes. As before, we begin with
$\mathbb{R}^N$:

\begin{df} A $0$-simplex in $\mathbb{R}^N$ is a point.
If $k>0$, a piecewise-analytic  $k$-simplex in $\mathbb{R}^N$  is a
semi-analytic set $S$ such that there is a homeomorphism
$\phi:\mathbb{R}^N\rightarrow  \mathbb{R}^N$ which maps $S$ and the
completion $\bar{S}$ onto the following sets
\begin{align}
\phi(S)\ =\ \{ (x^1,...,x^N)\in \mathbb{R}^N\ &:\ \sum_{i=1}^k x^k
<1, {\rm and}
\ x^{k+1},...,x^N=0 \}\\
\phi(\bar{S})\ =\ \{ (x^1,...,x^N)\in \mathbb{R}^N\ &:\ \sum_{i=1}^k
x^k \le 1, {\rm and} \ x^{k+1},...,x^N=0 \}.
\end{align}
\end{df}

And next we proceed with a manifold:
\begin{df}
A  $k$-simplex in a piecewise-analytic manifold is a subset
$S\subset M$ such that there is a piecewise-analytic chard
$(\chi,U)$ which maps $S$ onto a semi-analytic $k$-simplex in
$\mathbb{R}^N$.
\end{df}
The simplexes are used for partitions called triangulations:
\begin{df} A piecewise-analytic triangulation of subsets
$X_1,...,X_n\subset M$, where $M$ is a piecewise-analytic manifold,
is a family of pairwise disjoint subsets $S_1,...,S_m\subset M$,
such that $S_i$ is a $k_i$-simplex in $M$ for $i=1,...,n$, and
\begin{equation}
X_i\ =\ \bigcup_{k=1}^{n_i} S_{j_{ik}}, \ \ \ i=1,...,n,\ \  1\le
j_{ik}\le m.
\end{equation}
\end{df}

The property of the semi-analytic sets and piecewise-analytic
triangulations crucial for our work is \cite{semianalytic} (see
Lojasiewicz, page 463, Theorem 1,2):
\begin{prope}\label{trian}
Every finite family of piecewise-analytic simplexes in a piecewise
analytic manifold $M$ admits a piecewise-analytic triangulation.
\end{prope}
\medskip

Let us remark on the status of the framework we are using. The
theory of the semi-analytic sets is well established in the
mathematical literature \cite{semianalytic}. We apply the deep
results of that theory. There seems to be no unique generalization
of the notion of semi-analyticity to  a category of `semi-analytic'
manifolds, though. (Probably each generalization has its drawbacks
and limitations.) In the current paper we introduce  the weakest
from the point of view of our aims definition of a manifold
structure compatible with the definition of the semi-analytic sets .
It relies on a somewhat implicit (but not empty) definition of the
piecewise-analytic diffeomorphisms in $\mathbb{R}^N$. In \cite{lost}
we introduced a family of explicitely defined `semi-analytic'
diffeomorphsims, and used them to define `semi-analytic' manifolds.
They form a subclass in the class of the piecewise-analytic
manifolds defined above. Finally, our definitions should be also
compared with similar ideas of \cite{fleisch1}.

\medskip

%***************************************************
\noindent{\bf Acknowledgments} We have benefited from discussions
with  Abhay Ashtekar, John Baez, Marcin Bobie\'nski, Witold
Marciszewski, Tadeusz Mostowski, Hanno Sahlmann, Lee Smolin, Thomas
Thiemann and Andrzej Trautman. The work was partially supported by
the Polish Ministerstwo Nauki i Informatyzacji grant 1 P03A 015 29.
%***************************************************
%***************************************************

\end{document}